\documentclass[aps,prb,amsmath,amssymb,twocolumn]{revtex4-2}

\usepackage{amsmath, amssymb, bbold}
\usepackage[usenames,dvipsnames]{xcolor}
\usepackage{subfig}
\usepackage{hyperref}
\hypersetup{
	colorlinks,
	citecolor=RoyalBlue,
	filecolor=black,
	linkcolor=ForestGreen,
}

\usepackage{comment}

\usepackage{graphicx}
\usepackage{tikz,lipsum,lmodern}

\graphicspath{{./}}

\begin{document}
	
\title{%
Work Sum Rule for Open Quantum Systems
}

\author{Parth Kumar} \email[Corresponding author: ]{parthk@arizona.edu}
\author{Caleb M. Webb} \email{calebmwebb@arizona.edu}
\author{Charles A. Stafford} \email{stafford@physics.arizona.edu}
\affiliation{Department of Physics, University of Arizona, 1118 East Fourth Street, Tucson, Arizona 85721, USA}
\date{\today}
	
\begin{abstract}
A key question in the thermodynamics of open quantum systems is how to partition thermodynamic quantities such as entropy, work, and internal energy between the system and its environment.  We show that the only partition under which entropy is non-singular is based on a partition of Hilbert-space, which assigns half the system-environment coupling to the system and half to the environment.  However, quantum work partitions non-trivially under Hilbert-space partition, and we derive a Work Sum Rule %
that accounts for %
quantum work at a distance.  
All state functions of the system are shown to be path independent once this nonlocal quantum work is properly accounted for.
Our %
results are illustrated with application to 
a driven resonant level %
strongly coupled to a reservoir. 

\end{abstract}

\maketitle

The program of scaling the laws of thermodynamics down to the nanoscale and beyond has proven to be exceptionally challenging.
While substantial progress has been made \cite{michelQuantumThermodynamicsEmergence2009, binderThermodynamicsQuantumRegime2018,10.1088/2053-2571/ab21c6,strasbergQuantumStochasticThermodynamics2022,campisiColloquiumQuantumFluctuation2011,jarzynskiEqualitiesInequalitiesIrreversibility2011,seifertStochasticThermodynamicsFluctuation2012,shiThermalProbingEnergy2009,kimUltrahighVacuumScanning2012,Lee2013,neumannHighprecisionNanoscaleTemperature2013,mecklenburgNanoscaleTemperatureMapping2015,mengesTemperatureMappingOperating2016,cuiQuantizedThermalTransport2017,Mosso2017,Shastry2020_STTh,Toyabi2010_Szilard_exp,berutExperimentalVerificationLandauer2012,koskiExperimentalRealizationSzilard2014,devoretQuantumMachinesMeasurement2014a,liuPeriodicallyDrivenQuantum2021}, there exists little consensus %
on some of the most foundational questions, especially in the extreme quantum limit. %
Formulating the First Law of Thermodynamics for open quantum systems begs the question of how to partition quantum observables between the system and its environment.  Various competing schemes have been proposed in the literature \cite{ludovicoDynamicalEnergyTransfer2014,espositoNatureHeatStrongly2015,bruchQuantumThermodynamicsDriven2016,haughianQuantumThermodynamicsResonantlevel2018,talknerColloquiumStatisticalMechanics2020,
strasbergFirstSecondLaw2021a,lacerdaQuantumThermodynamicsFast2022,bergmannGreenFunctionPerspective2021}, many of which involve assigning part of the interfacial energy to the system and the remainder to the environment.  

In this Letter, we analyze the thermodynamics of a quasi-statically driven open quantum system strongly coupled to its environment, and
show that entropy is only well-defined under a Hilbert-space partition, which divides the system-environment coupling equally between the system and the environment.  
Internal energy, chemical work, and entropy are partitioned straightforwardly under a partition of Hilbert space.
However, the partition of quantum work is non-trivial in this framework.  We therefore derive a Work Sum Rule %
that accounts for the thermodynamic effect of %
quantum work at a distance.  
All state functions of the system are shown to be path independent once this nonlocal quantum work is properly accounted for.

The Internal Energy $U(t)$ of the universe (consisting of system + environment) with Hamiltonian $H(t)$ is %
\begin{equation}\label{eqn_integy_gen_def}
    U(t):=\langle H(t) \rangle\,,
\end{equation}
where $\langle\,\rangle$ denotes the quantum statistical average $\langle H(t) \rangle=\mathrm{Tr}\{H(t)\rho(t)\}$, where $\rho(t)$ is the density matrix of the universe at time $t$ and $\mathrm{Tr}\{\}$ denotes the trace over the full Fock space. The (inclusive) rate of Work done by external forces on the universe is  \cite{jarzynskiComparisonFarfromequilibriumWork2007,campisiColloquiumQuantumFluctuation2011,talknerColloquiumStatisticalMechanics2020}
\begin{equation}\label{eqn_totalextpower_def}
  \dot{W}_{ext}(t):= \frac{d}{dt}\langle H(t)\rangle= \langle \dot{H}(t) \rangle\,,
\end{equation}
where the second equality follows from the von Neumann equation for the density matrix (see Appendix \ref{app_totextpower_id}%
). 

In this Letter, we consider a time-dependent quantum universe of fermions without inter-particle interactions. In that case, 
the internal energy and power delivered can be expressed in terms of the single-particle Green's function
$G^{<}(t,t')$ as
\begin{equation}\label{eqn_intyegy_NEGF_gen}
    U(t)=-i\mathbb{Tr}\{h(t)G^{<}(t,t)\} \,,
\end{equation}
and 
\begin{equation}\label{eqn_extpower_NEGF_gen}
  \dot{W}_{ext}(t)= -i\mathbb{Tr}\{\dot{h}(t)G^{<}(t,t)\} \,,
\end{equation}
respectively, where $h(t)=h_S(t)+h_R +h_{SR}(t)$ is the matrix representation of $H(t)$ in the 1-body Hilbert-space, and $\mathbb{Tr}$ denotes a trace over this Hilbert-space 
(see Appendix \ref{app_GF_wigner_vars_defs}
for the definition and evaluation of $G^{<}$). %
Here $h_S(t)$ is the driven Hamiltonian of the quantum subsystem of interest, $h_R$ is the Hamiltonian of the macroscopic reservoir, and $h_{SR}(t)$ is the
coupling Hamiltonian.

We consider a quantum system that can exchange energy and particles with a reservoir in equilibrium at temperature $T$ and chemical potential $\mu$. 
The system is driven quasi-statically so that it remains in equilibrium with the reservoir throughout the driving protocol. Under these conditions, the equal-time Green's function takes the quasi-equilibrium form
\begin{equation}
 G^{<(0)}(t,t)=\int d\epsilon f(\epsilon)A^{(0)}(t,\epsilon)\,,
 \label{eq:G<0}
\end{equation}
where $A^{(0)}(t,\epsilon)=\delta(\epsilon-h(t))$ is the spectral function in the quasi-static limit (see Appendix \ref{app_GF_wigner_vars_defs})
and $f(\epsilon)=(1+e^{\beta(\epsilon-\mu)})^{-1}$ is the Fermi-Dirac distribution. Here the superscript $A^{(n)}$ denotes the order in time derivatives of the driving Hamiltonian.  We note that Eq.\ \eqref{eq:G<0} is exact to all orders in the system-reservoir coupling $h_{SR}(t)$, but omits terms involving $\dot{h}(t)$, $\ddot{h}(t)$, etc.

The relevant thermodynamic quantities for such a universe are evaluated as follows (see Appendix \ref{app_thermo_dos_spaecfn_rel_dervs}). 
The internal energy is given by
\begin{equation}\label{eqn_qstat_integy_dos_rel}
 U^{(0)}(t)=\int d\epsilon \, g(t,\epsilon)f(\epsilon)\epsilon \,,
\end{equation} 
where $g(t,\epsilon)=\mathbb{Tr}\{A^{(0)}(t,\epsilon)\}$ is the quasi-static density of states. %
Similarly, the quasi-static power delivered may be computed using Eqs.\ \eqref{eqn_extpower_NEGF_gen} and \eqref{eq:G<0} as 
\begin{equation}\label{eqn_qstat_extpower_spec_rel}
    \dot{W}_{ext}^{(1)}(t)=\int d\epsilon\,f(\epsilon)\mathbb{Tr}\{\dot{h}(t)A^{(0)}(t,\epsilon)\}\,.
\end{equation}
The quasi-static entropy is given by 
\begin{equation}\label{eqn_totalentropy_dos_rel}
  \langle S(t) \rangle %
  = S^{(0)}(t)=\int d\epsilon \,g(t,\epsilon)s(\epsilon)\,,  
\end{equation} 
where the entropy operator $S(t)=-\ln\rho(t)$, $s(\epsilon)=\beta(\epsilon-\mu)f(\epsilon)+\ln(1+e^{-\beta(\epsilon-\mu)})$, %
and we set $k_B=1$.
The mean number of particles  is given by
\begin{equation}\label{eqn_totalpartnum_dos_rel}
 N^{(0)}(t)=\int d\epsilon \, g(t,\epsilon)f(\epsilon) \,.  
\end{equation}
The grand canonical potential %
of the universe is %
$\Omega^{(0)}(t)=U^{(0)}(t)-TS^{(0)}(t)-\mu N^{(0)}(t)$,
which can be expressed as
\begin{equation} \label{eqn_totalgrandpot_dos_rel}
 \Omega^{(0)}(t)=\int d\epsilon \, g(t,\epsilon)\omega(\epsilon)\,,
\end{equation}
where $\omega(\epsilon)=-\frac{1}{\beta}\ln(1+e^{-\beta(\epsilon-\mu)})$.

The first variations of the thermodynamic quantities satisfy %
    $\dot{\Omega}^{(1)}(t)=\dot{U}^{(1)}(t)-T\dot{S}^{(1)}(t)-\mu \dot{N}^{(1)}(t)$.
Importantly, the first variation of $\Omega$ is equal to the external work %
\begin{equation}\label{eqn_omega_work_global_rel}
    \dot{W}_{ext}^{(1)}(t)=\dot{\Omega}^{(1)}(t)
\end{equation}
(see Appendices \ref{appsubsec_global_fdmt_thermo_id_derv} and \ref{app_tot_extwork_omega_equality_derv}
for derivations). %

A central quantity in scanning probe microscopy is the \textit{local density of states} (LDOS) %
\cite{%
binnigScanningTunnelingMicroscopy1983,binnigAtomicForceMicroscope1986,kalininScanningProbeMicroscopy2007,chenIntroductionScanningTunneling2021,muraltScanningTunnelingPotentiometry1986,williamsScanningThermalProfiler1986,shiThermalProbingEnergy2009,kimQuantitativeMeasurementScanning2011,yuHighresolutionSpatialMapping2011,kimUltraHighVacuumScanning2012a,mengesQuantitativeThermometryNanoscale2012,kimUltrahighVacuumScanning2012,Lee2013,mecklenburgNanoscaleTemperatureMapping2015,mengesTemperatureMappingOperating2016,staffordLocalTemperatureInteracting2016,shastryTemperatureVoltageMeasurement2016,Shastry2020_STTh}, which provides a firm experimental basis to construct local thermodynamic state functions in real quantum systems.
The LDOS of subsystem $\gamma$ is defined as \cite{dattaQuantumTransportAtom2005,moskaletsScatteringMatrixApproach2011} %
   $g_{\gamma}(t,\epsilon):= %
   \mathbb{Tr}\{\Pi_\gamma \delta(\epsilon-h(t))\}$,
where $\Pi_{\gamma}=\int_{x\in\gamma}\,dx |x\rangle\langle x|$ is the projection operator onto subspace $\gamma$ of the single-particle Hilbert-space.
$g_{\gamma}(t,\epsilon)$ %
gives the \textit{local spectrum} of the nonlocal operator $h(t)$.

The partitioned thermodynamic quantities $U_\gamma(t)$, $S_\gamma(t)$, $N_\gamma(t)$, and $\Omega_\gamma(t)$ %
are defined \cite{staffordLocalEntropyNonequilibrium2017,shastryThirdLawThermodynamics2019,shastryTheoryThermodynamicMeasurements2019}
by simply replacing $g(t,\epsilon)$ by $g_\gamma(t,\epsilon)$ in Eqs.\ \eqref{eqn_qstat_integy_dos_rel}, \eqref{eqn_totalentropy_dos_rel}, \eqref{eqn_totalpartnum_dos_rel}, and \eqref{eqn_totalgrandpot_dos_rel},
and are the quantum statistical averages of partitioned quantum observables $H|_\gamma$, $S|_\gamma$, and $N|_\gamma$, respectively, where %
$O|_\gamma$ is the Fock-space operator corresponding to the following operator defined on the single-particle Hilbert-space \cite{webbHowPartitionQuantum2024} (see Appendix \ref{app_hilbertspace_part_derv})%

\begin{equation}\label{eqn_hilb_sp_partition_def}
o|_{\gamma}=\frac{1}{2}\{\Pi_{\gamma},o\}\,,
\end{equation} 
where $o$ is the single-particle Hilbert-space operator corresponding to the global Fock-space operator $O=\sum_\gamma O|_\gamma$, and the anticommutator 
(defined as $\{a,b\}=ab+ba$) ensures the hermiticity of $O|_\gamma$.

In analogy with Eq.\ \eqref{eqn_omega_work_global_rel}, we define the {\it rate of thermodynamic work done} on subsystem $\gamma$ as
$\dot{W}_\gamma^{(1)}(t):=\dot{\Omega}_{\gamma}^{(1)}(t)$,
leading to the First Law of Thermodynamics for a quantum subsystem %
    $\dot{U}_{\gamma}^{(1)}(t)=T\dot{S}_{\gamma}^{(1)}(t)+\mu \dot{N}_{\gamma}^{(1)}(t)+\dot{W}_\gamma^{(1)}(t)$.
However, in general, %
$\dot{W}^{(1)}_\gamma(t)\neq \langle \dot{H}|_\gamma(t)\rangle$, so that the rate of thermodynamic work done on a given subsystem 
{\it is not equal} to the expectation value of the power operator partitioned on that subsystem. 
Instead, %
\begin{equation}\label{eqn_omegacurr_def}
\dot{W}^{(1)}_{\gamma}(t) :=\dot{\Omega}_{\gamma}^{(1)}(t)
=\langle \dot{H}|_{\gamma}(t)\rangle+I^W_\gamma(t)\,, 
\end{equation}
where $I^W_{\gamma}(t)$ represents the instantaneous quantum flow of free energy into subsystem $\gamma$ induced by the external drive $\dot{H}(t)$.
$I^W_{\gamma}(t)$ can be thought of as (the rate of) quantum work at a distance. %
Quantum work is inherently nonlocal because even if the external drive is local, the quantum states acted upon are nonlocal.
This nonlocal work predicted in driven quantum systems is reminiscent of the phenomena of measurement-induced energy teleportation \cite{hottaProtocolQuantumEnergy2008,ikedaDemonstrationQuantumEnergy2023} 
or conditional work at a distance
\cite{elouardInteractionFreeQuantumMeasurementDriven2020} in autonomous quantum systems.

The Hamiltonian of the open quantum system is 
    $H(t)=H_S(t)+H_R+H_{SR}(t)$,
where
 $H_S(t)=\sum_{n,m}[h_{S}(t)]_{nm}d^{\dagger}_{n}d_{m}$
is the system Hamiltonian and
$H_R=\sum_{k}\epsilon_{k}c_{k}^{\dagger}c_{k}$
is the reservoir Hamiltonian.
Finally, the coupling Hamiltonian $H_{SR}(t)$, describing the interface between the system and the reservoir, is 
$H_{SR}(t)=\sum_{k,n}[V_{kn}(t)c_{k}^{\dagger}d_{n}+\mbox{h.c.}]$.
Definition \eqref{eqn_hilb_sp_partition_def} implies that the Hamiltonian partitioned on the system is
\begin{equation}
    H|_S(t)=H_{S}(t)+\frac{1}{2}H_{SR}(t),
\end{equation}
so that the coupling Hamiltonian is partitioned equally between the system and the reservoir.

The following Sum Rule for external work %
can be derived for open systems
(see Appendix \ref{app_adiabatic_thermo_opendsys_dervs} 
for a derivation)
\begin{equation}\label{eqn_worksumrule_opensys}
   \dot{W}_{ext}^{(1)}(t)=
   \dot{\Omega}_S^{(1)}(t)+\delta\dot{\Omega}_R^{(1)}(t) \,,
\end{equation}  
where $\delta\Omega_R=\Omega_R-\Omega_{R,0}$
is the change in the reservoir grand potential induced due to the change in its spectrum as a result of its coupling to the system. A similar construction was introduced by Friedel \cite{friedelMetallicAlloys1958} to describe the cloud of screening charge induced in a metal due to the presence of an impurity.

An interesting and important case to consider is that where only the system is time-dependent so that $\dot{H}|_S(t)=\dot{H}_S(t)$ and $\dot{H}|_R(t)=0$.  Nonetheless, generically $\delta\dot{\Omega}_R(t)\neq 0$, so the work sum rule becomes
\begin{equation}\label{eqn_open_sum_rule_local}
   \langle \dot{H}_S(t)\rangle \stackrel{\dot{H}_{SR}=0}{=} \dot{\Omega}_S^{(1)}(t)+\delta\dot{\Omega}_R^{(1)}(t) \,.
\end{equation}
Thus, even if external forces act {\it only inside the system}, the instantaneous thermodynamic work done on the reservoir is nonzero.
For this case, Eqs.\ \eqref{eqn_omegacurr_def} and \eqref{eqn_open_sum_rule_local} imply that the rate of nonlocal quantum work done on the system is
$I^W_{S} \stackrel{\dot{H}_{SR}=0}{=}-\delta\dot{\Omega}^{(1)}_R$,  
or minus the rate of nonlocal quantum work done on the reservoir. %

Using %
the internal energy $U_S(t)=\langle H|_S(t)\rangle$, the 1st Law for the open system becomes 
\begin{equation}\label{eqn_firstlaw_oqs}
\frac{d}{dt}\langle H_{S}(t)+\frac{1}{2}H_{SR}(t)\rangle = T\dot{S}_S^{(1)}+\mu\dot{N}_S^{(1)} +\dot{W}_S^{(1)} , 
\end{equation}
where $S_S$ is the entropy partitioned on the system Hilbert-space%
, and 
\begin{equation}\label{eqn_W_S_def}
    \dot{W}_S^{(1)}=\dot{W}_{ext}^{(1)}-\delta \dot{\Omega}_R^{(1)}=\langle \dot{H}|_S \rangle + I^W_S
\end{equation}
is the rate of thermodynamic work done on the system.

Using the NEGF formalism \cite{%
haugQuantumKineticsTransport2007,rammerQuantumFieldTheory2007,stefanucciNonequilibriumManyBodyTheory2013}, each term in the First Law [Eq.\ \eqref{eqn_firstlaw_oqs}] can be expressed in terms of the quasi-static system Green's functions (see Appendix \ref{app_adiabatic_thermo_opendsys_dervs}
for derivations). The terms contributing to the (partitioned) quasi-static power [Eq.\ \eqref{eqn_qstat_extpower_spec_rel}] delivered by external forces are  
\begin{equation}\label{eqn_qstat_extpower_S_NEGF}
   \langle \dot{H}_S(t) \rangle
   = 
   \int \frac{d\epsilon}{\pi}\,f(\epsilon)\,
   \mathbb{ImTr}\Bigg\{\dot{h}_S(t) \mathcal{G}^{A(0)}(t,\epsilon)  
    \Bigg\}\,,
\end{equation} 
\begin{equation}\label{eqn_qstat_extpower_SR_NEGF}
   \langle \dot{H}_{SR}(t)\rangle
   = 
   \int \frac{d\epsilon}{\pi}\,f(\epsilon)\,\mathbb{ImTr}\Bigg\{\dot{\Sigma}^{A}(t,\epsilon)\mathcal{G}^{A(0)}(t,\epsilon) 
    \Bigg\} \,, 
\end{equation} 
where $\mathcal{G}^{A(0)}$ and $\Sigma^{A}$ are the quasi-static advanced system Green's function and self-energy, respectively, defined in Appendix \ref{app_GF_wigner_vars}.
The rate of nonlocal work done on the system is 
\begin{eqnarray}\label{eqn_nonlocal_qwork_sys}
    I_S^W(t) = -\frac{1}{2}\langle \dot{H}_{SR}(t)\rangle &+&\int \frac{d\epsilon}{\pi}\, \mathbb{ImTr}\Bigg\{\frac{\partial \mathcal{G}^{A(0)}(t,\epsilon)}{\partial t}\frac{\partial \Sigma^{A}(t,\epsilon)}{\partial\epsilon}\nonumber \\ &-&\frac{\partial \mathcal{G}^{A(0)}(t,\epsilon)}{\partial\epsilon}\frac{\partial \Sigma^{A}(t,\epsilon)}{\partial t}\Bigg\}\omega(\epsilon)\,,
\end{eqnarray}
where the second term on the RHS may be interpreted as an instantaneous flow of free energy into the system induced by the time-dependent external drive, while the nonlocality of the first term on the RHS is trivial since $H_{SR}$ is itself nonlocal.

We note that in the broad-band limit [$\partial_\epsilon \Sigma^{A}(t,\epsilon)=0$] the rate of nonlocal quantum work vanishes, in contrast to the notion proposed in Ref.\ \cite{haughianQuantumThermodynamicsResonantlevel2018} and discussed in Ref.\ \cite{bergmannGreenFunctionPerspective2021}(see Appendix \ref{app_bblimit_nonlocal_qwork}).

\begin{figure}
\centering
\includegraphics[width=8cm,height=10cm]{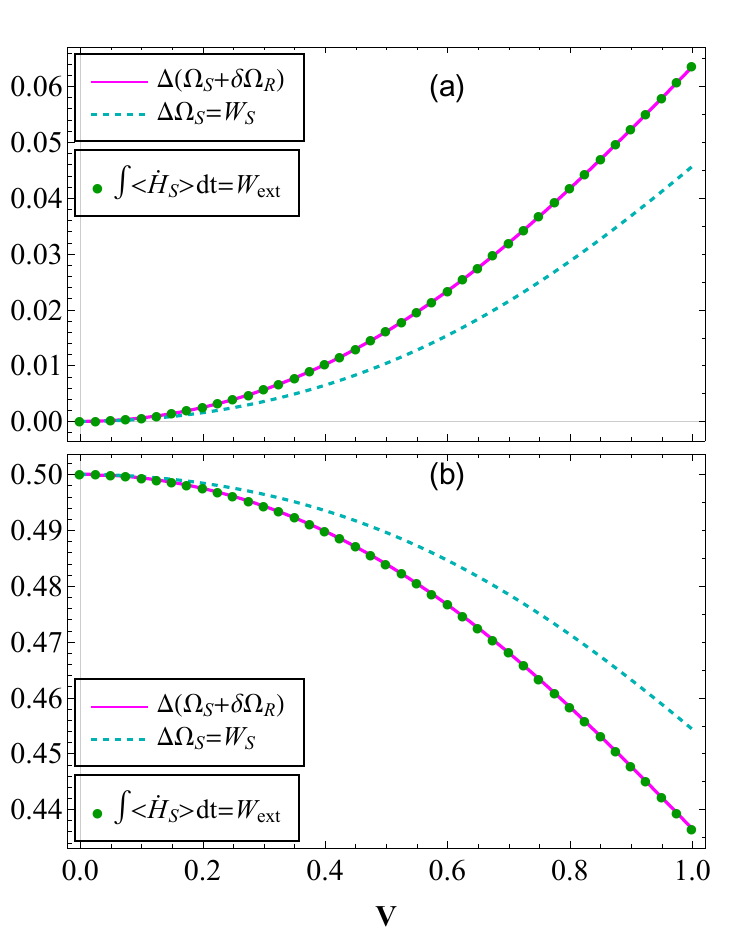} 
\captionsetup{justification=raggedright, singlelinecheck=false}	
\caption{%
Verification of the Work Sum rule [Eqs.\ \eqref{eqn_worksumrule_opensys}, \eqref{eqn_open_sum_rule_local}] for
an open quantum system: the resonant-level model. %
Only the level $\varepsilon_s(t)$ is driven (a) from 1 to 1.5, and (b) from -1.5 to -1. The reservoir is maintained at temperature $T=0.02$ and chemical potential $\mu=0$, with  %
$t_0=1.25$.
}
\label{fig_rlm_sumrule_erdriven}
\end{figure}

The work sum rule and Hilbert-space partition of the thermodynamics derived above are illustrated with an application to the driven resonant-level model. (See Appendix \ref{app_mult_resv}
for an extension to multiple reservoirs and analysis of a driven two-level system coupled to two reservoirs.)
The system Hamiltonian is %
$ %
    H_{S}(t)=\varepsilon_{s}(t)d^{\dagger}d$,
the reservoir is modeled as a semi-infinite tight-binding chain with hopping integral $t_0$,
$ %
    H_{R}= %
    t_0\sum_{j=1}^\infty (c_{j}^{\dagger}c_{j+1}+\mbox{h.c.}),
$ %
and the interface between system and reservoir is modelled by 
$ %
H_{SR}(t)=V(t)d^{\dagger}c_{1}+\mbox{h.c.}$
Two driving protocols are %
investigated:  In {\it protocol 1}, the level $\varepsilon_s(t)$ is varied %
while $V$ is held fixed.  In {\it protocol 2}, both
$\varepsilon_s(t)$ and $V(t)$ are varied.

A verification of the Work Sum Rule 
for protocol 1 [Eq.\ \eqref{eqn_open_sum_rule_local}]
is shown in Fig.\ \ref{fig_rlm_sumrule_erdriven}, where the nonlocal quantum work on the reservoir
is clearly visible as the difference between $\Delta \Omega_S$ (dashed blue curve) and the total work done $W_{\rm ext}$ (green dots). The nonlocal quantum work done on the reservoir may be positive [Fig.\ref{fig_rlm_sumrule_erdriven}(a)] or negative [Fig.\ref{fig_rlm_sumrule_erdriven}(b)] depending on system parameters, and increases in magnitude as the coupling $V$ between the system and reservoir increases.  %

Let us compare our analysis of the thermodynamics of a quasi-statically driven open quantum system with some previous frameworks found in the literature
\cite{ludovicoDynamicalEnergyTransfer2014,espositoNatureHeatStrongly2015,bruchQuantumThermodynamicsDriven2016,haughianQuantumThermodynamicsResonantlevel2018,strasbergFirstSecondLaw2021a}. These frameworks can be described as $\alpha$-partitions of the internal energy, where $\alpha \in [0,1]$ describes the fraction of the coupling Hamiltonian $H_{SR}$ included in the internal energy of the open system.
In the $\alpha$-partition, the internal energy of the open quantum system is
\footnote{Ref.\ \cite{espositoNatureHeatStrongly2015} uses the convention $\alpha\rightarrow 1-\alpha$.}
\begin{equation}\label{eqn_U_alpha}
    \alpha\mbox{-}U_{S}(t)=\langle H_S(t)+\alpha H_{SR}(t)\rangle\,,
\end{equation}
and the rate of external work done on the system is identified by some authors as \cite{espositoNatureHeatStrongly2015}
\begin{equation}\label{eqn_W_alpha}
   \alpha\mbox{-}\dot{W}_{S}(t)=
   \langle \dot{H}_S(t)+\alpha \dot{H}_{SR}(t)\rangle\,. 
\end{equation}
The Hilbert-space partition proposed in this Letter corresponds to setting $\alpha=1/2$ in Eq.\ \eqref{eqn_U_alpha}. However, as discussed above, the rate of
external work done on the system is given by Eq.\ \eqref{eqn_W_S_def}, and cannot in general be expressed as in Eq.\ \eqref{eqn_W_alpha} for any value of $\alpha$.

For independent quantum particles, the entropy operator $-\ln\hat{\rho}^{(0)}$ under quasi-static driving is also a one-body observable, and can be partitioned in the same way
\footnote{The particle number $N_S$ that appears in the chemical work term in the First Law [Eq.\ \eqref{eqn_firstlaw_oqs}] is typically independent of $\alpha$.}.
One finds for the $\alpha$-partition of the entropy of subsystem $\gamma$ (see Appendix \ref{app:alpha_S})
\begin{eqnarray}\label{eqn_entropy_part_open}
    \alpha\mbox{-}S^{(0)}_\gamma(t) %
    &=& 2\alpha \int d\epsilon \, %
    g_\gamma (t,\epsilon) s(\epsilon) \nonumber \\ &+&  (1-2\alpha)\int d\epsilon\int d\epsilon'\,\mathbb{Tr}\{\Tilde{A}_{\gamma}(t,\epsilon)\Tilde{A}_{\gamma}(t,\epsilon')\}\nonumber\\ &\times&  [-f(\epsilon)\ln(f(\epsilon'))-(1-f(\epsilon))\ln(1-f(\epsilon'))]\,, \nonumber\\ 
\end{eqnarray}
where $\Tilde{A}_{\gamma}(t,\epsilon)=\Pi_{\gamma}\delta(\epsilon-h(t))\Pi_{\gamma}$. 

Although it might appear %
that one could construct the partitioned thermodynamics for arbitrary values of $\alpha$, a severe problem arises in
$\alpha$-$S_\gamma$ if $\alpha \neq 1/2$.  For $\alpha > 1/2$, the term beginning on the second line %
Eq.\ \eqref{eqn_entropy_part_open} is negative and unbounded, while for $\alpha < 1/2$, the partitioned entropy of the subspace complementary to $\gamma$ is negative and unbounded. {\it Since entropy is a non-negative quantity, any partition that yields a negative subsystem entropy should be ruled out on principle as unphysical.}  Moreover, in fermionic systems there are generically tightly-bound core states with occupancy $f\rightarrow 1$ and high-lying scattering states with occupancy $f\rightarrow 0$,
for both of which %
Eq.\ \eqref{eqn_entropy_part_open} is undefined for $\alpha \neq 1/2$.  Thus we are forced to conclude that the only physically allowable partitioning of the entropy is $\alpha=1/2$, namely, the Hilbert-space partition %
as previously proposed in Ref.\ \cite{shastryThirdLawThermodynamics2019}. %

\begin{figure}%
	
\centering
 \includegraphics[width=8cm,height=10cm]{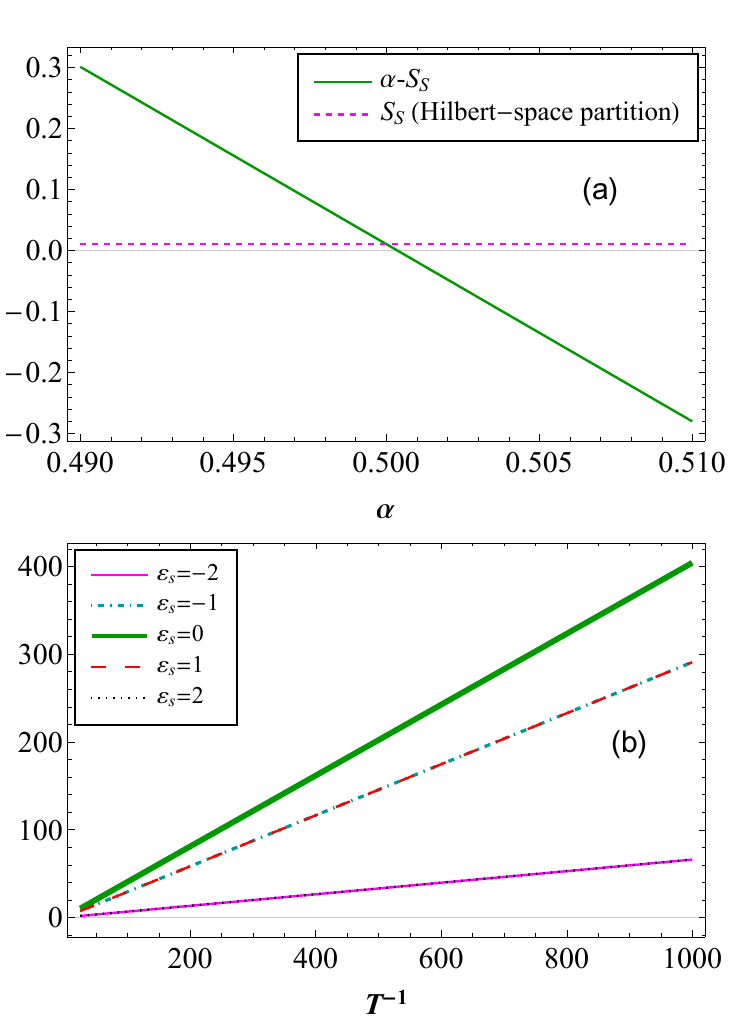}
	
\captionsetup{justification=raggedright, singlelinecheck=false}	
\caption{%
Results for the $\alpha$-partition of the entropy %
[Eq.\ \eqref{eqn_entropy_part_open}]
in the resonant-level model with $\mu=0$, $V=1$, and $t_0=1.25$. 
(a) Comparison of $\alpha\mbox{-}S_S$ to the entropy under Hilbert-space partition ($\alpha=1/2$) at $T=0.02$. (b) Coefficient of ($1-2\alpha$) in Eq.\ \eqref{eqn_entropy_part_open} (2nd term on the RHS) versus inverse temperature %
for various values of the level energy $\varepsilon_s=0,\pm 1, \pm 2$. %
}
\label{fig_rlm_alphapartition_abs_entropy}
\end{figure}

Figure \ref{fig_rlm_alphapartition_abs_entropy}(a) shows a comparison of the small physical value of the system entropy for $\alpha=1/2$ and the large (positive or negative) unphysical entropy for $\alpha\neq 1/2$ for the resonant-level model with
$\varepsilon_s=-1$, $V=1$, $\mu=0$, and $T=0.02$. 
Under Hilbert-space partition, the system entropy %
is bounded by $0\leq S_S \leq \ln 2$.
Note that the slope of the green curve representing $\alpha\mbox{-}S_S$ tends to $-\infty$ as $T\rightarrow 0$, indicating a severe contradiction of the Third Law of Thermodynamics for $\alpha\neq 1/2$, as pointed out previously in Refs.\ \cite{espositoNatureHeatStrongly2015,shastryThirdLawThermodynamics2019}.
Figure \ref{fig_rlm_alphapartition_abs_entropy}(b) plots the coefficient of $(1-2\alpha)$ in Eq.\ \eqref{eqn_entropy_part_open} (second term on the RHS) as a function of inverse temperature at $\mu=0$ for several values of the resonant level $\varepsilon_s$, indicating that this unphysical contribution to the entropy partition diverges $\propto 1/T$ as $T\rightarrow 0$, as can be readily understood from Eq.\ \eqref{eqn_entropy_part_open} and the functional form of the Fermi-Dirac distribution.

\begin{figure}%
	
\centering
 \includegraphics[width=8cm,height=10cm]{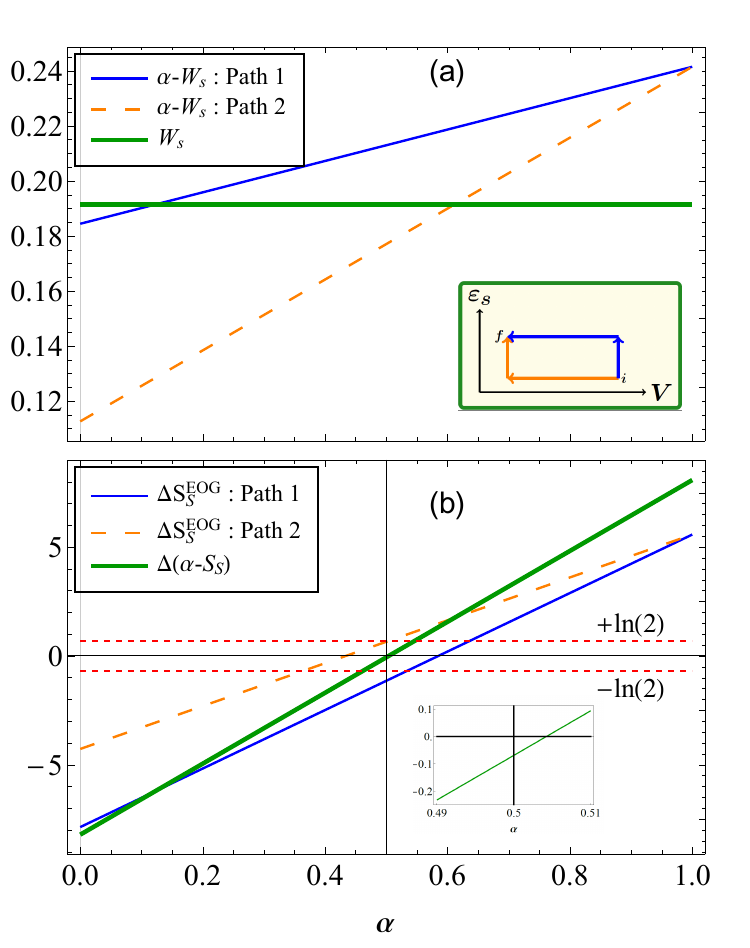}
	
\captionsetup{justification=raggedright, singlelinecheck=false}	
\caption{$\alpha$-Partition of the thermodynamics %
in the resonant-level model%
, where both the level $\varepsilon_s(t)$: $0\rightarrow 1$ and coupling $V(t)$: $0.6 \rightarrow 0.4$ are driven along two different paths [see protocols in inset (a)]. %
Here $T=0.02$, $\mu=0$, and $t_0=1.25$.
(a) Two definitions of work, $W_S$ [defined in Eq.\ \eqref{eqn_W_S_def}] and $\alpha\mbox{-}W_S$ [defined in Eq.\ \eqref{eqn_W_alpha}] %
versus $\alpha$.
(b) The change in the statistical mechanical partition of the entropy $\Delta (\alpha\mbox{-}S_{S})$ [Eq.\ \eqref{eqn_entropy_part_open}] and 
$\Delta S_S^{\rm EOG}$ [Eq.\ \eqref{eqn:S_EOG}], proposed in Ref.\ \cite{espositoNatureHeatStrongly2015}, versus $\alpha$ for the same processes.
The bound $|\Delta S_S| \leq \ln 2$ under Hilbert-space partition %
is shown as red dashed lines. 
}
\label{fig_rlm_alphapartition}
\end{figure}

Figure \ref{fig_rlm_alphapartition}(a) plots $W_S$ [defined in Eq.\ \eqref{eqn_W_S_def}] and $\alpha\mbox{-}W_S$ [defined in Eq.\ \eqref{eqn_W_alpha}] for two different paths in protocol 2 (see inset) as a 
function of $\alpha$. %
The quasi-static work for a system in the grand canonical ensemble at constant $T$ and $\mu$ should be independent of path.
It is clear that $\alpha\mbox{-}W_S$, the definition of the work done on the system proposed in Ref.\ \cite{espositoNatureHeatStrongly2015},
is only path independent in the limit $\alpha\rightarrow 1$, wherein 
$\lim_{\alpha\rightarrow 1}\alpha\mbox{-}W_S = W_{ext}$, which unsurprisingly is indeed path independent.  However, our definition [Eq.\ \eqref{eqn_W_S_def}]
of the work $W_S$ done on the Hilbert space of the system, including nonlocal quantum work, is path independent and generally not equal to the total external work $W_{ext}$ for nonzero system-reservoir coupling.

Using their definitions of internal energy [Eq.\ \eqref{eqn_U_alpha}] and of work done on the system [Eq.\ \eqref{eqn_W_alpha}], Esposito, Ochoa, and Galperin \cite{espositoNatureHeatStrongly2015} use the thermodynamic identity
\begin{equation}\label{eqn:S_EOG}
    T\Delta S_S^{\rm EOG} = \Delta (\alpha\mbox{-}U_S) -\mu \Delta N_S - \alpha\mbox{-}W_S
\end{equation}
to define the change in system entropy $\Delta S_S^{\rm EOG}$ for a process.  $\Delta S_S^{\rm EOG}$ is plotted in Fig.\ \ref{fig_rlm_alphapartition}(b) as a function of $\alpha$ for two different paths in protocol 2.
Also plotted is the change in $\alpha\mbox{-}S_S$, the $\alpha$-partition of statistical mechanical entropy [Eq.\ \eqref{eqn_entropy_part_open}].
Although our model is slightly different than that used in Ref.\ \cite{espositoNatureHeatStrongly2015} (we utilize a semi-infinite 1D tight-binding model of the reservoir while Ref.\ \cite{espositoNatureHeatStrongly2015} uses a phenomenological self-energy), the
quantitative results for the two models are comparable (see Appendix \ref{app_GF_wigner_vars_defs}).

As shown in Fig.\ \ref{fig_rlm_alphapartition}(b) and in Ref.\ \cite{espositoNatureHeatStrongly2015}, $\Delta S_S^{\rm EOG}$ is path independent only for $\alpha=1$.  Based on this fact, Esposito, Ochoa, and Galperin suggest that $\alpha=1$ should be chosen in the partition of the internal energy [Eq.\ \eqref{eqn_U_alpha}], despite the acknowledged violation of the Third Law of Thermodynamics for $\alpha=1$. 
It is noteworthy that the magnitude of $\Delta S_S^{\rm EOG}$ greatly exceeds $\ln 2$ [see red dashed lines in Fig.\ \ref{fig_rlm_alphapartition}(b)], the maximum entropy of the system under Hilbert-space partition.  In fact, $\Delta S_S^{\rm EOG}|_{\alpha=1}=5.57$ is opposite in sign and 80.9 times larger in magnitude than the actual entropy change $\Delta S_S=-0.068$ for the process shown in Fig.\ \ref{fig_rlm_alphapartition}.
We would argue that there is no room in the Hilbert space of the system for so much entropy, no matter how the level is broadened and shifted due to a finite coupling to the reservoir.

The problem with the entropy \cite{espositoNatureHeatStrongly2015} defined thermodynamically via Eq.\ \eqref{eqn:S_EOG}, which is not equal to the statistical mechanical partition $\alpha\mbox{-}S_S$ for any value of $\alpha$, stems from the incorrect definition [Eq.\ \eqref{eqn_W_alpha}] of the work done on the system, which does not take into account the nonlocal quantum work $\int I_S^W dt$.  Once the correct definition of work [Eq.\ \eqref{eqn_W_S_def}] is used, the 
Hilbert-space partition ($\alpha=1/2$) of the statistical mechanical entropy [Eq.\ \eqref{eqn_entropy_part_open}] satisfies the First Law [Eq.\ \eqref{eqn_firstlaw_oqs}], and is path independent, as shown in Fig.\ \ref{fig_rlm_alphapartition}(b).

In this Letter, we have derived a Work Sum Rule describing the thermodynamic effects of nonlocal quantum work.
An open quantum system is analogous to a quantum impurity problem, and nonlocal quantum work is a thermodynamic effect analogous to the screening charge that must be included in the Friedel sum rule \cite{friedelMetallicAlloys1958}.
We emphasize that our thermodynamic partition of the entropy is different than the usual information-theoretic partition 
based on the reduced state of a quantum subsystem \cite{breuer2007}; due to the nonlocality of quantum information, the reduced-state description 
introduces %
entanglement entropy
that is not associated with any thermodynamic process \cite{webbHowPartitionQuantum2024}. 

The thermodynamic partition proposed in this Letter is based on a partition of Hilbert space, and
is consistent with the analysis of Ref.\ \cite{ludovicoDynamicalEnergyTransfer2014}, which circumnavigates the issue of partitioning quantum work.
Alternative attempts \cite{espositoNatureHeatStrongly2015} %
to partition the thermodynamics of open quantum systems have failed due to an incorrect partition of quantum work.
\\

We thank Carter Eckel, Ferdinand Evers, and Yiheng Xu for insights developed during the preliminary stages of this work, and acknowledge useful discussions on the thermodynamics of the resonant-level model with Michael Galperin. 

\bibliography{%
arxivV2.bbl
}

\clearpage

\onecolumngrid

\appendix

\section{Derivation of Eq.\ 2} %
\label{app_totextpower_id}

Using the inclusive definition of external work \cite{jarzynskiComparisonFarfromequilibriumWork2007,campisiColloquiumQuantumFluctuation2011,talknerColloquiumStatisticalMechanics2020},
the quantum statistical average rate of external work done on a system with Hamiltonian $H(t)$ is defined to be
\begin{equation}
    \dot{W}_{ext}(t) := \frac{d}{dt} \langle H(t) \rangle,
    \nonumber
\end{equation}
where $H(t)$ is the total Hamiltonian describing the subsystem of interest, the environment, and the system-environment coupling.
It is straightforward to show that the mean rate of external work is equal to the expectation value of the total power operator
[second equality in Eq.\ \eqref{eqn_totalextpower_def}]: %
\begin{equation}\label{eqn_totalextpower_result}
    \dot{W}_{ext}(t)=\langle\dot{H}(t)\rangle\,. 
\end{equation}
This can be shown by first noting that since $\langle H(t)\rangle=\mathrm{Tr}\{\rho(t)H(t)\}$, it follows from the product rule that
\begin{equation}
 \frac{d}{dt}\langle H(t)\rangle=\mathrm{Tr}\{\dot{\rho}(t)H(t)+\rho(t)\dot{H}(t)\} \,. 
\end{equation}
Eq.\ \eqref{eqn_totalextpower_result} then follows from inserting the von Neumann equation of motion for the density matrix
\begin{equation}
    i\hbar\dot{\rho}(t)=[H(t),\rho(t)]
\end{equation}
in the above equation and using the cyclicity of the Trace.

\section{Nonequilibrium Green's functions in the quasi-static limit}\label{app_GF_wigner_vars}

\subsection{Key definitions and relations}\label{app_GF_wigner_vars_defs}

The general system lesser two-time Green's function is defined as 
\begin{equation}
    G_{nm}^{<}(t,t')=i\langle d_{m}^{\dagger}(t')d_{n}(t)\rangle\,.
\end{equation}
The advanced system, tunneling, and reservoir Green's function are defined, respectively, as
\begin{equation}
    G^A_{nm}(t,t')=i\theta(t'-t)\langle\{d_n(t),d^\dagger_m(t')\}\rangle\,,
\end{equation}
\begin{equation}
    G^A_{nk}(t,t')=i\theta(t'-t)\langle\{d_n(t),c^\dagger_k(t')\}\rangle\,,
\end{equation}
and 
\begin{equation}
    G^A_{kk'}(t,t')=i\theta(t'-t)\langle\{c_k(t),c^\dagger_{k'}(t')\}\rangle\,,
\end{equation}
where $\theta(x)$ is the Heaviside step function.

The analysis of quasi-static driving is facilitated by performing the so-called Wigner transformation \cite{haugQuantumKineticsTransport2007,rammerQuantumFieldTheory2007} %
of the time-dependent quantities defined by the substitutions 
\begin{equation}
    \tau=t-t' \,,
\end{equation}
and
\begin{equation}
 \bar{t}=\frac{t+t'}{2}\,,   
\end{equation}
where $\tau$ is the fast dynamical quantum time-scale while $\bar{t}$ is the slow time-scale of the quasi-static drive. %
We denote the Wigner transformed lesser Green's function as $\widetilde{G}(\bar{t},\tau)$, and its %
Fourier transform with respect to the quantum time $\tau$ is defined as 
\begin{equation}
\mathcal{G}(\bar{t},\epsilon):=\int d\tau\, e^{i\epsilon\tau}\widetilde{G}(\bar{t},\tau)\,,
\end{equation}
with the inverse transform is given by 
\begin{equation}
\widetilde{G}(\bar{t},\tau)=\int \frac{d\epsilon}{2\pi}\, e^{-i\epsilon\tau}\mathcal{G}(\bar{t},\epsilon) \,.
\end{equation}
The equal-time Green's function is obtained by setting $\tau=0$ and $\bar{t}=t$, so that
\begin{equation}
G^<_{nm}(t,t)=\widetilde{G}_{nm}^{<}(\bar{t}=t,0)= \int \frac{d\epsilon}{2\pi}\,\mathcal{G}_{nm}^{<}(\bar{t}=t,\epsilon) \,.   
\end{equation}

Derivatives of observables with respect to the quasi-static timescale $\bar{t}$ are assumed to be infinitessimal, and
a series expansion of the Green's function in powers of $\partial/\partial\bar{t}$ can be developed \cite{haugQuantumKineticsTransport2007}
\begin{equation}
    \widetilde{G}(\bar{t},\tau)= \widetilde{G}^{(0)}(\bar{t},\tau)+\widetilde{G}^{(1)}(\bar{t},\tau)+\widetilde{G}^{(2)}(\bar{t},\tau)+...\,, 
\end{equation}
where the superscript $(n)$ denotes the power of $\partial /\partial \bar{t}$.
Only the lowest-order term in the above expansion is needed to describe 
quasi-static driving. 
In particular, 
\begin{equation}
    \mathcal{G}^{<(0)}(\bar{t},\epsilon)=2\pi i f(\epsilon)A^{(0)}(\bar{t},\epsilon),
\end{equation}
where the quasi-static spectral function is 
$A^{(0)}(\bar{t},\epsilon)=\delta(\epsilon-h(\bar{t}))$.  

The submatrix of $A^{(0)}$ within the system Hilbert-space may be expressed as
\begin{equation}\label{eqn_A_quasi}
    A^{(0)}(\bar{t},\epsilon)=\frac{1}{2\pi i}[\mathcal{G}^{A(0)}(\bar{t},\epsilon)-\mathcal{G}^{R(0)}(\bar{t},\epsilon)]\, ,
\end{equation}
where $\mathcal{G}^{A(0)}$ and $\mathcal{G}^{R(0)}$ are the Wigner-transformed advanced and retarded system Green's functions, respectively, in the quasi-static limit
\begin{equation}
    \mathcal{G}^{A/R(0)}(\bar{t},\epsilon)=\left[1\epsilon -h(\bar{t}) - \Sigma^{A/R}(\bar{t},\epsilon)\right]^{-1},
\end{equation}
and the (advanced) self-energy in the quasi-static limit is %
\begin{equation}\label{eqn_adv_self_egy}
 \Sigma^{A}_{nm}(\bar{t},\epsilon)
 =\sum_kV_{kn}(\bar{t})\mathcal{G}^{A(0)}_{kk}(\epsilon)V^{*}_{km}(\bar{t}) \,. 
\end{equation} 

\begin{figure}
    \centering
   \includegraphics{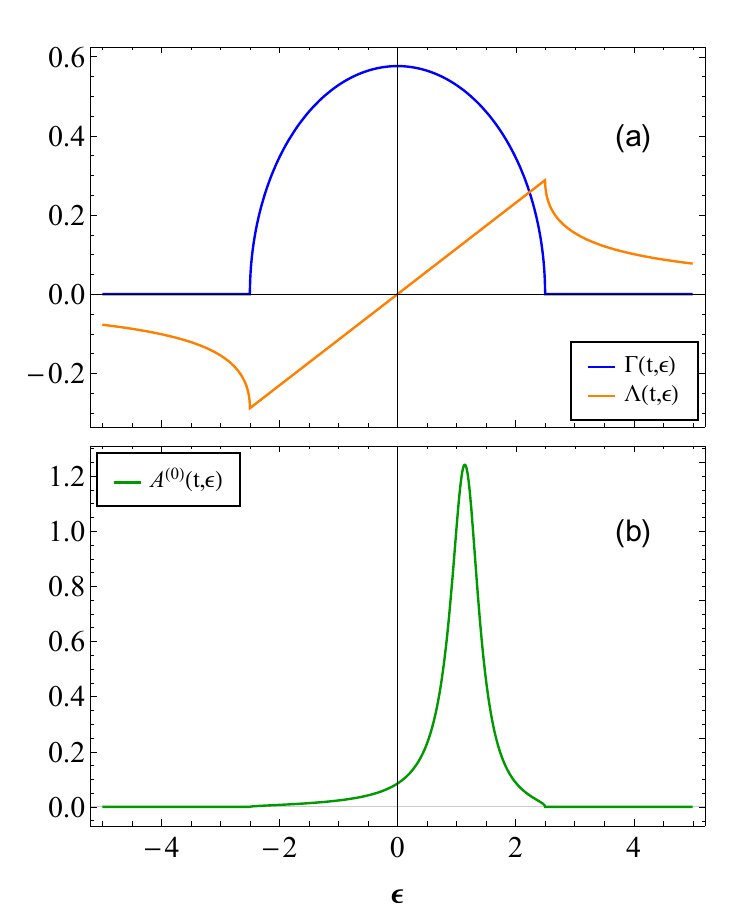}
    \captionsetup{justification=raggedright, singlelinecheck=false}
    \caption{(a) The real and imaginary parts of the advanced self-energy $\Sigma^A(t,\epsilon)=\Lambda(t,\epsilon) + i\Gamma(t,\epsilon)/2$, and (b) the spectral function $A^{(0)}(t,\epsilon)$ at $\varepsilon_s(t)=1$, for the resonant-level model as a function of energy for $V(t)=0.6$ and $t_0=1.25$. %
    }
    \label{fig_self_egy}
\end{figure}

For the resonant-level model considered in this Letter, 
\begin{equation}
    \Sigma^{A}(\bar{t},\epsilon)=\Lambda(\bar{t},\epsilon)+i\frac{\Gamma(\bar{t},\epsilon)}{2}
\end{equation}
is simply a complex number whose real and imaginary parts are shown in Fig.\ \ref{fig_self_egy} and may be calculated explicitly as \cite{dattaChapterLevelBroadening2005} %

\begin{equation}
\Lambda(\bar{t},\epsilon)=\frac{V^2(\bar{t})}{2t_0^2}
\begin{cases}
(\epsilon-\epsilon_0)+[(\epsilon-\epsilon_0)^2-4t_0^2)]^{\frac{1}{2}},  & \epsilon-\epsilon_0 < -2t_0, \\
(\epsilon-\epsilon_0),  & |\epsilon-\epsilon_0|<2t_0, \\
(\epsilon-\epsilon_0)-[(\epsilon-\epsilon_0)^2-4t_0^2)]^{\frac{1}{2}},  & \epsilon-\epsilon_0 > 2t_0,
\end{cases}
\end{equation}
\begin{equation}
\Gamma(\bar{t},\epsilon)= 
\begin{cases}
\frac{2V^2(\bar{t})}{t_0}\Big(1-\Big(\frac{\epsilon-\epsilon_0}{2t_0}\Big)^{2}\Big)^{\frac{1}{2}}, & |\epsilon-\epsilon_0|<2t_0, \\
0, & \text{otherwise}.
\end{cases}
\label{tbchain_Gamma_eqn}
\end{equation}

\subsection{Relation of thermodynamic quantities to density of states [Eqs.\ \eqref{eqn_qstat_integy_dos_rel}, \eqref{eqn_qstat_extpower_spec_rel}, and \eqref{eqn_totalpartnum_dos_rel}]}
\label{app_thermo_dos_spaecfn_rel_dervs}

From Eqs.\ \eqref{eqn_intyegy_NEGF_gen} and \eqref{eq:G<0} %
it follows that 
\begin{equation}
    U^{(0)}(t)=\int d\epsilon \, f(\epsilon)\mathbb{Tr}\{h(t)A^{(0)}(t,\epsilon)\}\,.
\end{equation}
From the identity $h(t)A^{(0)}(t,\epsilon)=\epsilon A^{(0)}(t,\epsilon)$ (from the spectral theorem), and the fact that the quasi-static density of states $g(t,\epsilon)=\mathbb{Tr}\{A^{(0)}(t,\epsilon)\}$, Eq.\ \eqref{eqn_qstat_integy_dos_rel} follows immediately. The analogous expressions for quasi-static power [Eq.\ \eqref{eqn_qstat_extpower_spec_rel}], entropy [Eq.\ \eqref{eqn_totalentropy_dos_rel}], and particle number [Eq.\  \eqref{eqn_totalpartnum_dos_rel}] can be derived similarly. See Appendix \ref{app:alpha_S} for further details on the entropy.

\section{Global Quasi-static dynamics and thermodynamics
}
\label{app_adiabatic_thermo_global_dervs}

The quantum adiabatic theorem \cite{Sakurai2011}
tells us that the state of a system with an adiabatically varying Hamiltonian $h(t)$ is given by a linear combination of vectors of the form 
\begin{equation}
    |\psi_{\nu}(t)\rangle=e^{i\theta_{\nu}(t)}e^{i\gamma_{\nu}(t)}|\nu(t)\rangle\,,
\end{equation}
where $\theta_{\nu}(t)=-\frac{1}{\hbar}\int_{0}^{t}dt'\,\epsilon_{\nu}(t')$ is the so-called dynamical phase,  $\gamma_{\nu}(t)=i\int_{0}^{t}dt'\,\langle\psi_{\nu}(t')|\frac{\partial}{\partial t'}\psi_{\nu}(t')\rangle$ is the so-called geometrical phase, and $|\nu(t)\rangle$ solves the instantaneous Hamiltonian eigenvalue equation $h(t)|\nu(t)\rangle=\epsilon_{\nu}(t)|\nu(t)\rangle$. 

\subsection{Derivation of the Global Fundamental Thermodynamic Identity}
\label{appsubsec_global_fdmt_thermo_id_derv}

Using $g(t,\epsilon)=\sum_\nu \delta(\epsilon-\epsilon_\nu(t))$, where $\epsilon_\nu(t)$ are the instantaneous eigenvalues of %
$h(t)$, and it is understood that the sum becomes an integral for continuous spectra,
the thermodynamic quantities [Eqs.\ \eqref{eqn_qstat_integy_dos_rel}, \eqref{eqn_totalentropy_dos_rel}, \eqref{eqn_totalpartnum_dos_rel}, and \eqref{eqn_totalgrandpot_dos_rel}], may be written
\begin{equation}
    U(t)=\sum_{\nu}f_{\nu}(t)\epsilon_{\nu}(t)\,,
\end{equation}
\begin{equation}
    S(t)=-\sum_{\nu} \left\{  f_{\nu}(t)\ln f_{\nu}(t)+  (1- f_{\nu}(t))\ln[1-f_{\nu}(t)] \right\} \,,
\end{equation}
\begin{equation}
    N(t)=\sum_{\nu}f_{\nu}(t)\,,
\end{equation}
and
\begin{equation}
    \Omega(t)=-\frac{1}{\beta}\sum_{\nu}\ln[1+e^{-\beta(\epsilon_{\nu}(t)-\mu)}]\,,
\end{equation}
 where $f_{\nu}(t)\equiv f(\epsilon_{\nu}(t))$.  
 
It follows by the chain rule that the first variations can be computed as 
\begin{equation}
    \dot{U}^{(1)}(t)=\sum_{\nu}(\dot{f}_{\nu}(t)\epsilon_{\nu}(t)+f_{\nu}(t)\dot{\epsilon}_{\nu}(t))\,,
\end{equation}
\begin{equation}
    \dot{S}^{(1)}(t)=\beta\sum_{\nu}(\epsilon_{\nu}(t)-\mu)\dot{f}_{\nu}(t)\,,
\end{equation}
\begin{equation}
    \dot{N}^{(1)}(t)=\sum_{\nu}\dot{f}_{\nu}(t)\,,
\end{equation}
and
\begin{equation}\label{eqn_totgrandpot_firstvar_egbasis}
    \dot{\Omega}^{(1)}(t)=\sum_{\nu}f_{\nu}(t)\dot{\epsilon}_{\nu}(t)\,.
\end{equation}
The global fundamental thermodynamic identity %
 $\dot{\Omega}^{(1)}(t)=\dot{U}^{(1)}(t)-T\dot{S}^{(1)}(t)-\mu \dot{N}^{(1)}(t)$
follows straightforwardly from the addition and rearrangement of the last four equations.

\subsection{Derivation of Eq.\ \eqref{eqn_omega_work_global_rel}}
\label{app_tot_extwork_omega_equality_derv}
 Eq.\ \eqref{eqn_totalextpower_result} can be written in the instantaneous eigenbasis of the Hamiltonian as 
\begin{equation}
    \dot{W}_{ext}(t)=\sum_{\nu}f(\epsilon_{\nu}(t))\langle\nu(t)|\dot{h}(t)|\nu(t)\rangle\,.
\end{equation}
The instantaneous eigenstates and eigenvalues of $h(t)$ satisfy \cite{Sakurai2011}
\begin{equation}
    \langle\nu(t)|\dot{h}(t)|\mu(t)\rangle=\dot{\epsilon}_{\nu}(t)\delta_{\mu\nu}+(\epsilon_{\mu}(t)-\epsilon_{\nu}(t))\langle\nu(t)|\dot{\mu}(t)\rangle\,,
\end{equation}
using which we may write 
\begin{equation}
    \dot{W}_{ext}(t)=\sum_{\nu}f(\epsilon_{\nu}(t))\dot{\epsilon}_{\nu}(t)\,,
\end{equation}
where the RHS is identical to that of Eq.\ \eqref{eqn_totgrandpot_firstvar_egbasis}, thus establishing Eq.\ \eqref{eqn_omega_work_global_rel}.

\section{Motivation for the operator Hilbert-space partition [Eq.\ 12]}\label{app_hilbertspace_part_derv}

We begin by noting that the LDOS %
can be rewritten as
\begin{equation}\label{eqn_LDOS-proj-spec_rel}
    g_{\gamma}(t,\epsilon):=\mathbb{Tr}\{\Pi_{\gamma}A^{(0)}(t,\epsilon)\}
    =\int_{x\in \gamma} dx \, \langle x | \delta (\epsilon-h(t)|x\rangle\,,
\end{equation}
which makes clear that the LDOS is simply the integral of the local spectrum over subspace $\gamma$.

To motivate the Hilbert-space partition %
of a quantum observable \cite{webbHowPartitionQuantum2024}, let us consider a dynamical observable $O$ that is compatible with the Hamiltonian $H(t)$.
Then the corresponding Hilbert-space operator 
$o$ commutes with $h(t)$, 
and $o|\epsilon\rangle =o(\epsilon)|\epsilon\rangle$, where $h(t)|\epsilon\rangle=\epsilon|\epsilon\rangle$ and $o(\epsilon)$ is an eigenvalue of $o$.
In analogy with the definitions of $U_\gamma$ and $N_\gamma$, %
define
\begin{equation}\label{eqn_O_gamma_compat}
    O_\gamma^{(0)}(t) := \int d\epsilon \, g_\gamma (t,\epsilon) f(\epsilon)o(\epsilon)
    =\int d\epsilon \, \mathbb{Tr}\{\Pi_{\gamma}
    \delta(\epsilon-h(t)) %
    \} f(\epsilon)o(\epsilon).
\end{equation}
Because $[o,h(t)]=0$, the operator $o$ can be brought inside the trace so that
\begin{equation}
    O_\gamma^{(0)}(t) = \int d\epsilon \, \mathbb{Tr}\{\Pi_{\gamma}
    \delta(\epsilon-h(t)) o\} f(\epsilon).
\end{equation}
Finally, from the cyclicity of the trace and the the compatibility of $o$ and $h(t)$, this can be expressed using Eq.\ \eqref{eqn_hilb_sp_partition_def} as
\begin{equation}\label{eqn_O_gamma_gen}
    O_\gamma^{(0)}(t) = \int d\epsilon \, \mathbb{Tr}\{o|_{\gamma} \,
    \delta(\epsilon-h(t))\} f(\epsilon) = \langle O|_\gamma(t)\rangle ^{(0)}.
\end{equation}
This completes the demonstration that the partition of thermodynamic quantities based on the local spectrum (LDOS) %
corresponds to the quantum statistical average of the partitioned observables defined by Eq.\ \eqref{eqn_hilb_sp_partition_def}.  This
equivalence holds for dynamical observables compatible with the Hamiltonian.  For the application to statistical observables, such as entropy, see Appendix \ref{app:alpha_S}.  For dynamical observables that do not commute with the Hamiltonian, such as the power operator $\dot{H}(t)$, Eq.\ \eqref{eqn_hilb_sp_partition_def} is taken as the definition of the partitioned observable, but the quantum statistical average is no longer given by 
Eq.\ \eqref{eqn_O_gamma_compat}, but is instead given by the more general Eq.\ \eqref{eqn_O_gamma_gen}.

\section{Quasi-static dynamics and thermodynamics of an open quantum system}\label{app_adiabatic_thermo_opendsys_dervs}

To prove the work sum rule for open systems [Eq.\ \eqref{eqn_worksumrule_opensys}], we start by evaluating the power delivered by the external forces 
[Eq.\ \eqref{eqn_totalextpower_def}]. For quasi-static driving, Eq.\ \eqref{eqn_qstat_extpower_spec_rel} gives
\begin{equation}\label{eqn_W_ext_app_def}
    \dot{W}_{ext}(t)=\int \frac{d\epsilon}{\pi}\,f(\epsilon)\,\mathbb{ImTr}\{\dot{h}(t)\mathcal{G}^{A (0)}(t,\epsilon)\}\,,
\end{equation}
where we have used Eq.\ \eqref{eqn_A_quasi}. %
Treating the system and coupling terms in $\dot{h}(t)=\dot{h}_S(t)+ \dot{h}_{SR}(t)$ separately, and using the equation of motion for the coupling Green's function,
$\mathcal{G}^{A(0)}_{nk}(t,\epsilon)=\sum_m \mathcal{G}^{A(0)}_{nm}(t,\epsilon) V^{*}_{km}(t) \mathcal{G}^{A}_{kk}(\epsilon)$,
Eq.\ \eqref{eqn_W_ext_app_def} can be written entirely in terms of the system Green's function as
\begin{equation}\label{eqn_W_ext_app_result}
   \dot{W}_{ext}(t)=\int \frac{d\epsilon}{\pi}\,f(\epsilon)\,\mathbb{ImTr}\Bigg\{(\dot{h}_S(t) + \dot{\Sigma}^{A}(t,\epsilon))\mathcal{G}^{A (0)}(t,\epsilon)  
    \Bigg\}\,.
\end{equation}
where $\Sigma^{A}_{nm}(t,\epsilon)=\sum_kV_{kn}(t)\mathcal{G}^{A}_{kk}(\epsilon)V^{*}_{km}(t)$ is the advanced coupling self-energy. %

To compute the time derivative of the grand canonical potential, we may write, in analogy with the Friedel Sum Rule \cite{friedelMetallicAlloys1958} %
\begin{equation}\label{eqn_grandpot_NEGF}
\Omega^{(0)}_S(t)+\delta\Omega^{(0)}_R(t)=\int_{-\infty}^{\infty} \,\frac{d\epsilon}{\pi} \mathbb{ImTr}\Bigg\{\Bigg(1-\frac{\partial \Sigma^{A}(t,\epsilon)}{\partial \epsilon}\Bigg)\mathcal{G}^{A(0)}(t,\epsilon)\Bigg\}\omega(\epsilon)  \,.
\end{equation}
The first variation of this is
\begin{equation}\label{eqn_grandpot_NEGF_firstvar}
   \dot{\Omega}^{(1)}(t)=\int_{-\infty}^{\infty} \,\frac{d\epsilon}{\pi} \mathbb{ImTr}\Bigg\{\Bigg(-\frac{\partial^2 \Sigma^{A}(t,\epsilon)}{\partial t \partial \epsilon}\Bigg)\mathcal{G}^{A(0)}(t,\epsilon)+\Bigg(1-\frac{\partial \Sigma^{A}(t,\epsilon)}{\partial \epsilon}\Bigg)\frac{\partial \mathcal{G}^{A(0)}(t,\epsilon)}{\partial t}\Bigg\}\omega(\epsilon)  \,, 
\end{equation}
which gives the NEGF results of Eqs.\ \eqref{eqn_qstat_sysgrandpot_NEGF} and \eqref{eqn_qstat_resgrandpot_NEGF}. 

To prove the work sum rule, we will show that Eqs.\ \eqref{eqn_W_ext_app_result} and \eqref{eqn_grandpot_NEGF_firstvar} are equal.
First, $\mathcal{G}^{A(0)}$ obeys the equation of motion
\begin{equation}\label{eqn_GA_eq_t}
  \frac{\partial \mathcal{G}^{A(0)}(t,\epsilon)}{\partial t}=  \mathcal{G}^{A(0)}(t,\epsilon)[\dot{h}_S(t)+\dot{\Sigma}^{A}(t,\epsilon)]\mathcal{G}^{A(0)}(t,\epsilon) \,,
\end{equation}
which follows from the identity
\begin{equation}\label{eqn_inv_mat_derv_id}
    \frac{dM^{-1}(x)}{dx}=-M^{-1}(x)\frac{dM(x)}{dx}M^{-1}(x)\,,
\end{equation}
where $M(x)$ is an invertible matrix which is a function of a real scalar $x$.
Using Eq.\ \eqref{eqn_GA_eq_t} in the second term on the RHS of the Eq.\ \eqref{eqn_grandpot_NEGF_firstvar}, along with the cyclicity of the trace, gives
\begin{equation}\label{eqn_grandpot_NEGF_firstvar_2}
   \dot{\Omega}^{(1)}(t)=\int_{-\infty}^{\infty} \,\frac{d\epsilon}{\pi} \mathbb{ImTr}\Bigg\{\Bigg(-\frac{\partial^2 \Sigma^{A}(t,\epsilon)}{\partial t \partial \epsilon}\Bigg)\mathcal{G}^{A(0)}(t,\epsilon)+\mathcal{G}^{A(0)}(t,\epsilon)\Bigg(1-\frac{\partial \Sigma^{A}(t,\epsilon)}{\partial \epsilon}\Bigg)\mathcal{G}^{A(0)}(t,\epsilon)[\dot{h}_S(t)+\dot{\Sigma}^{A}(t,\epsilon)]\Bigg\}\omega(\epsilon)  \,. 
\end{equation} 
Finally, using the identity  %
\begin{equation}
   \frac{\partial \mathcal{G}^{A(0)}(t,\epsilon)}{\partial \epsilon}= -\mathcal{G}^{A(0)}(t,\epsilon)\Bigg(1-\frac{\partial \Sigma^{A}(t,\epsilon)}{\partial \epsilon}\Bigg)\mathcal{G}^{A(0)}(t,\epsilon) %
\end{equation}
[obtained using Eq.\ \eqref{eqn_inv_mat_derv_id}]
in Eq.\ \eqref{eqn_grandpot_NEGF_firstvar_2}, we obtain  
\begin{equation}\label{eqn_grandpot_NEGF_firstvar_3}
   \dot{\Omega}^{(1)}(t)=\int_{-\infty}^{\infty} \,\frac{d\epsilon}{\pi} \mathbb{ImTr}\Bigg\{\Bigg(-\frac{\partial^2 \Sigma^{A}(t,\epsilon)}{\partial t \partial \epsilon}\Bigg)\mathcal{G}^{A(0)}(t,\epsilon)-\frac{\partial \mathcal{G}^{A(0)}(t,\epsilon)}{\partial \epsilon}[\dot{h}_S(t)+\dot{\Sigma}^{A}(t,\epsilon)]\Bigg\}\omega(\epsilon)  \,. 
\end{equation}
Performing an integration by parts of the second term on the RHS of the above equation, we get 
\begin{eqnarray}
  \int_{-\infty}^{\infty} \,\frac{d\epsilon}{\pi} \mathbb{ImTr}\Bigg\{\frac{\partial \mathcal{G}^{A(0)}(t,\epsilon)}{\partial \epsilon}[\dot{h}_S(t)+\dot{\Sigma}^{A}(t,\epsilon)]\Bigg\}\omega(\epsilon) &=& 
  -\int_{-\infty}^{\infty} \,\frac{d\epsilon}{\pi}\, f(\epsilon)
  \mathbb{ImTr}\Bigg\{\mathcal{G}^{A(0)}(t,\epsilon)[\dot{h}_S(t)+\dot{\Sigma}^{A}(t,\epsilon)]\Bigg\} \nonumber \\  &&
  -\int_{-\infty}^{\infty} \,\frac{d\epsilon}{\pi} \,\mathbb{ImTr}\Bigg\{\Bigg(\frac{\partial^2 \Sigma^{A}(t,\epsilon)}{\partial t \partial \epsilon}\Bigg)\mathcal{G}^{A(0)}(t,\epsilon)\Bigg\}\omega(\epsilon) \,, \nonumber  \\
\end{eqnarray}
where
the boundary term vanishes and we have used $\partial\omega(\epsilon)/\partial\epsilon=f(\epsilon)$.
Substituting this back into Eq.\ \eqref{eqn_grandpot_NEGF_firstvar_3}, 
we recover the result on the RHS of Eq.\ \eqref{eqn_W_ext_app_result},
thus proving the work sum rule, $\dot{W}_{ext}^{(1)}= \dot{\Omega}_S^{(1)}+\delta\dot{\Omega}_R^{(1)}$. In doing this we have also obtained the partitioned quasi-static rates of change of grand potential %
\begin{equation}\label{eqn_qstat_sysgrandpot_NEGF}
   \dot{\Omega}_S^{(1)}=\int_{-\infty}^{\infty} \,\frac{d\epsilon}{\pi} \mathbb{ImTr}\left\{
    \frac{\partial \mathcal{G}^{A(0)}(t,\epsilon)}{\partial t}
   \partial_t \mathcal{G}^{A(0)}(t,\epsilon)
  \right\}\omega(\epsilon) \, , 
\end{equation}
and  
 \begin{eqnarray}\label{eqn_qstat_resgrandpot_NEGF}
 \dot{\Omega}_R^{(1)}=&-&\int_{-\infty}^{\infty} \,\frac{d\epsilon}{\pi} \mathbb{ImTr}\Bigg\{\frac{\partial\Sigma^{A}(t,\epsilon)}{\partial \epsilon}\frac{\partial \mathcal{G}^{A(0)}(t,\epsilon)}{\partial t}\Bigg\}\omega(\epsilon)\nonumber \\ 
  &-&\int_{-\infty}^{\infty} \,\frac{d\epsilon}{\pi} \mathbb{ImTr}\Bigg\{\frac{\partial^2\Sigma^{A}(t,\epsilon)}{\partial t\partial \epsilon} \mathcal{G}^{A(0)}(t,\epsilon)\Bigg\}\omega(\epsilon)\,. \nonumber \\
 \end{eqnarray}
The NEGF expression for nonlocal work done on the system $I_S^W(t)$ [Eq.\ \eqref{eqn_nonlocal_qwork_sys}] can be obtained in a similar way.

 \begin{figure}
    \centering
    \includegraphics[width=0.45\textwidth,height=19cm]{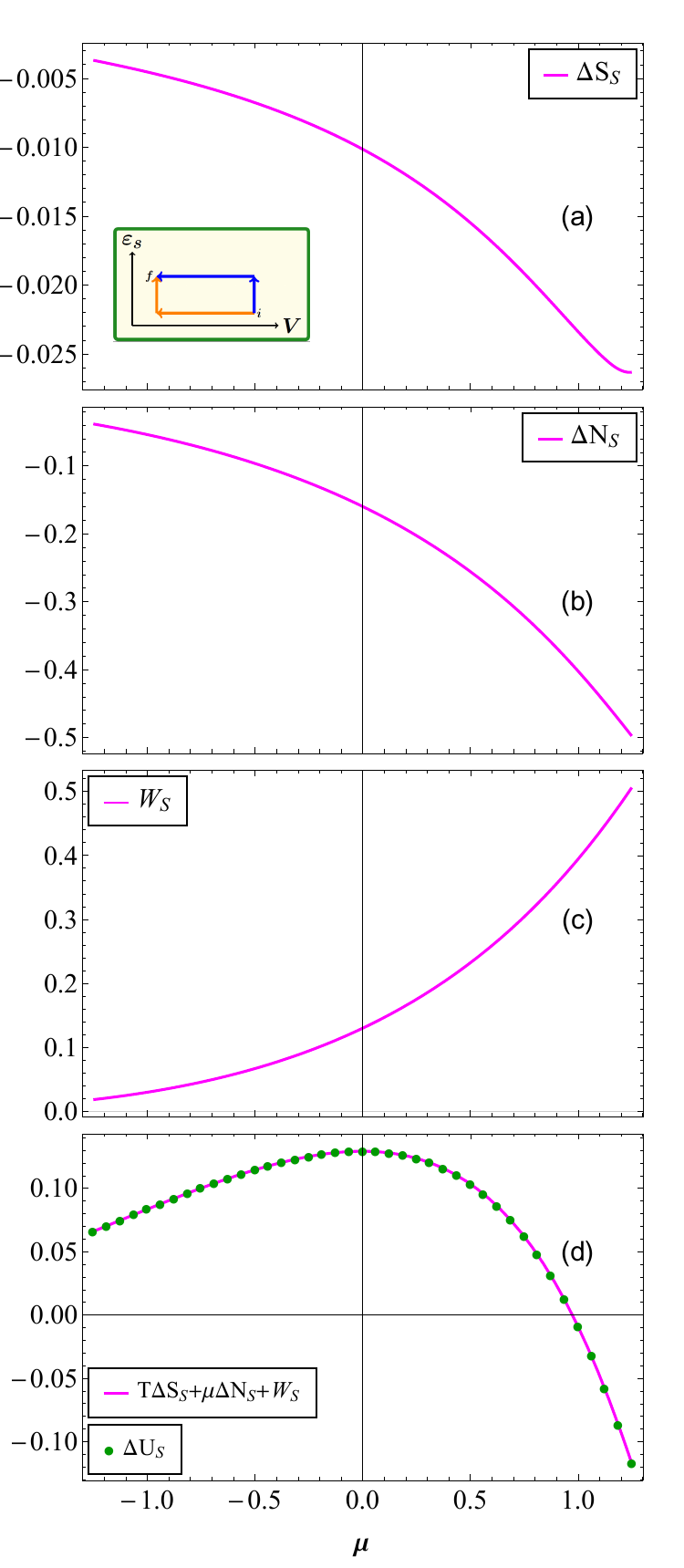}
    \captionsetup{justification=raggedright, singlelinecheck=false}
    \caption{The integrated terms (a) entropy $\Delta S_S$, (b) occupancy $\Delta N_S$, and (c) work $W_S$
    appearing in the 
First Law, Eq.\ \eqref{eqn_firstlaw_oqs}, %
in the resonant-level model, where both the 
level $\varepsilon_s(t)$ and the coupling $V(t)$ to the reservoir are driven. (d) Verification of the equality of the LHS and RHS of Eq.\ \eqref{eqn_firstlaw_oqs}.  Here the reservoir $T=0.02$, while the Hamiltonian parameters are varied along the two paths shown in the inset with $\varepsilon_s$: $0\rightarrow 1$ and $V$: $0.6\rightarrow 0.4$.}
\label{fig_rlm_EsVdriven_firstlaw_grid}
\end{figure}

In an analogous manner, the NEGF expressions for internal energy, entropy, and particle currents can be derived. We simply state the results below. %
\begin{equation}\label{eqn_qstat_sysintegy_NEGF}
     \dot{U}_S^{(1)}=\int_{-\infty}^{\infty} \,\frac{d\epsilon}{\pi} \mathbb{ImTr}\Bigg\{\frac{\partial \mathcal{G}^{A(0)}(t,\epsilon)}{\partial t}\Bigg\}f(\epsilon)\epsilon \, , 
\end{equation}
 \begin{eqnarray}\label{eqn_qstat_resintegy_NEGF}
   \delta \dot{U}_R^{(1)}=&-&\int_{-\infty}^{\infty} \,\frac{d\epsilon}{\pi} \mathbb{ImTr}\Bigg\{\Bigg(\frac{\partial\Sigma^{A}(t,\epsilon)}{\partial \epsilon}\Bigg)\Bigg(\frac{\partial \mathcal{G}^{A(0)}(t,\epsilon)}{\partial t}\Bigg)\Bigg\}f(\epsilon)\epsilon\nonumber \\ 
   &-&\int_{-\infty}^{\infty} \,\frac{d\epsilon}{\pi} \mathbb{ImTr}\Bigg\{\frac{\partial^2\Sigma^{A}(t,\epsilon)}{\partial t\partial \epsilon} \mathcal{G}^{A(0)}(t,\epsilon)\Bigg\}f(\epsilon)\epsilon\,.
 \end{eqnarray}
 \begin{equation}\label{eqn_qstat_sysentropy_NEGF}
     \dot{S}_S^{(1)}=\int_{-\infty}^{\infty} \,\frac{d\epsilon}{\pi} \mathbb{ImTr}\Bigg\{\frac{\partial \mathcal{G}^{A(0)}(t,\epsilon)}{\partial t}\Bigg\}s(\epsilon) \, , 
\end{equation}
 \begin{eqnarray}\label{eqn_qstat_resentropy_NEGF}
   \delta \dot{S}_R^{(1)}=&-&\int_{-\infty}^{\infty} \,\frac{d\epsilon}{\pi} \mathbb{ImTr}\Bigg\{\Bigg(\frac{\partial\Sigma^{A}(t,\epsilon)}{\partial \epsilon}\Bigg)\Bigg(\frac{\partial \mathcal{G}^{A(0)}(t,\epsilon)}{\partial t}\Bigg)\Bigg\}s(\epsilon)\nonumber \\ 
   &-&\int_{-\infty}^{\infty} \,\frac{d\epsilon}{\pi} \mathbb{ImTr}\Bigg\{\frac{\partial^2\Sigma^{A}(t,\epsilon)}{\partial t\partial \epsilon} \mathcal{G}^{A(0)}(t,\epsilon)\Bigg\}s(\epsilon)\,.
 \end{eqnarray}
 \begin{equation}\label{eqn_qstat_syspartnum_NEGF}
     \dot{N}_S^{(1)}=\int_{-\infty}^{\infty} \,\frac{d\epsilon}{\pi} \mathbb{ImTr}\Bigg\{\frac{\partial \mathcal{G}^{A(0)}(t,\epsilon)}{\partial t}\Bigg\}f(\epsilon) \, , 
\end{equation}
 \begin{eqnarray}\label{eqn_qstat_respartnum_NEGF}
   \delta \dot{N}_R^{(1)}=&-&\int_{-\infty}^{\infty} \,\frac{d\epsilon}{\pi} \mathbb{ImTr}\Bigg\{\Bigg(\frac{\partial\Sigma^{A}(t,\epsilon)}{\partial \epsilon}\Bigg)\Bigg(\frac{\partial \mathcal{G}^{A(0)}(t,\epsilon)}{\partial t}\Bigg)\Bigg\}f(\epsilon)\nonumber \\ 
   &-&\int_{-\infty}^{\infty} \,\frac{d\epsilon}{\pi} \mathbb{ImTr}\Bigg\{\frac{\partial^2\Sigma^{A}(t,\epsilon)}{\partial t\partial \epsilon} \mathcal{G}^{A(0)}(t,\epsilon)\Bigg\}f(\epsilon)\,.
 \end{eqnarray}

All the above thermodynamic quantities are plotted for the quasi-statically driven resonant-level model for two different protocols in Fig.\ \ref{fig_rlm_EsVdriven_firstlaw_grid}.

\section{Broad-band limit for the rate of nonlocal quantum work}\label{app_bblimit_nonlocal_qwork}
The rate of nonlocal work done [Eq.\ \eqref{eqn_nonlocal_qwork_sys}] vanishes in the broad-band limit provided the limit is taken in a physically meaningful manner. We may write the first term on the RHS of Eq.\ \eqref{eqn_nonlocal_qwork_sys} using  Eq.\ \eqref{eqn_qstat_extpower_SR_NEGF} in the broad-band limit as (since $\Lambda\rightarrow 0$)
\begin{equation}
    \langle \dot{H}_{SR}\rangle=\int \frac{d\epsilon}{2\pi}\,f(\epsilon)\mathbb{Tr}[\dot{\Gamma}(t,\epsilon)\mathbb{Re}\{G^A(t,\epsilon)\}]\,.
\end{equation}
We note first that $\Gamma(t,\epsilon)$ given by Eq.\ \eqref{tbchain_Gamma_eqn} is bounded by
\begin{equation}
    \Gamma(t,\epsilon)\leq \frac{2V^2(t)}{t_0}\theta(t_0-|\epsilon-\epsilon_0|).
\end{equation}
It therefore follows in the limit $T\rightarrow 0$ that
\begin{equation}
    |\langle \dot{H}_{SR}\rangle| \leq \left|\frac{V\dot{V}}{\pi t_0}\ln\Bigg\{\frac{(\Gamma/2)^2+(\mu-\varepsilon_s)^2)}{(\Gamma/2)^2+(2t_0+\epsilon_0-\varepsilon_s)^2)}\Bigg\}\right|.
\end{equation}
The limiting behavior for large $t_0$ is
\begin{equation}
    \langle \dot{H}_{SR}\rangle \stackrel{t_0\rightarrow\infty}{\sim} -\frac{2V\dot{V}}{\pi t_0}\ln\Bigg\{\frac{2t_0}{[(\Gamma/2)^2+(\mu-\varepsilon_s)^2)]^{1/2}}\Bigg\}\,.
\end{equation}
From this functional form, it is evident that if the broadband limit $t_0\rightarrow \infty$ is taken while the coupling $V$ is kept constant, then  $\langle \dot{H}_{SR}\rangle \rightarrow 0$. However, if the broadband limit $t_0\rightarrow \infty$ is taken while the tunneling-width matrix element $\Gamma$ is kept constant (forcing $V\rightarrow\infty$), then $\langle \dot{H}_{SR}\rangle$ diverges logarithmically. The latter is clearly an unphysical scenario since in any real setup the system-reservoir coupling element $V$ will have a finite value. 

Finally, the second and third terms in Eq.\ \eqref{eqn_nonlocal_qwork_sys} can be shown to also vanish in the broad-band limit using manipulations similar to those used to obtain the NEGF expression for $\delta\dot{\Omega}_R(t)$ [Eq.\ \eqref{eqn_qstat_resgrandpot_NEGF}].

\section{$\alpha$-Partition of the Entropy}
\label{app:alpha_S}
In this appendix, Fock-space operators are denoted with hats. %
For a system of independent fermions in the Grand Canonical Ensemble driven quasi-statically, the density matrix has the instantaneous equilibrium form
\begin{equation}
    \hat{\rho}^{(0)}= \prod_\nu [f_\nu \psi_\nu^\dagger \psi_\nu + (1-f_\nu)\psi_\nu\psi_\nu^\dagger],
\end{equation}
where $\psi_\nu^\dagger$ creates a fermion in an eigenstate of the instantaneous Hamiltonian $h(t)$, $\psi_\nu^\dagger |0\rangle = |\nu(t)\rangle$, where $h(t)|\nu(t)\rangle=\varepsilon_\nu(t)|\nu(t)\rangle$ and $f_\nu :=f(\varepsilon_\nu(t))$.
For such a system, the entropy operator has the form
\begin{equation}
    \hat{S}^{(0)}=-\ln\hat{\rho}^{(0)}=\sum_\nu[-\ln f_\nu \psi_\nu^\dagger\psi_\nu-\ln(1-f_\nu)\psi_\nu \psi_\nu^\dagger],
\end{equation}
which consists of a particle-like piece $\hat{S}^n$ and a hole-like piece $\hat{S}^p$, whose corresponding operators on the single-particle Hilbert-space are
\begin{eqnarray}
    s^n & = & -\sum_\nu \ln f_\nu |\nu\rangle \langle \nu|,\\
    s^p & = & -\sum_\nu \ln(1-f_\nu) |\nu\rangle \langle \nu|.
\end{eqnarray}
These can be partitioned according to Eq.\ \eqref{eqn_hilb_sp_partition_def} to obtain the Hilbert-space partition of the entropy.

For the $\alpha$-partition of the entropy, we need to separate $s=s^n+s^p$ into diagonal and off-diagonal pieces
\begin{equation}
    s_{SS} := \Pi_S (s^n + s^p) \Pi_S
\end{equation}
and
\begin{equation}
    s_{SR} := \Pi_S (s^n + s^p)\Pi_R + \Pi_R (s^n + s^p)\Pi_S.
\end{equation}
The $\alpha$-partition corresponds to the Hilbert-space operator
\begin{equation}
    s_{SS}+\alpha s_{SR} = (1-2\alpha) \Pi_S s \Pi_S + \alpha \{\Pi_S, s\},
\end{equation}
where we have used $1=2\alpha + (1-2\alpha)$ and $\Pi_S+\Pi_R=1$.  Clearly, for $\alpha=1/2$, $s_{SS}+\alpha s_{SR}=s|_S$, as defined by Eq.\ \eqref{eqn_hilb_sp_partition_def}.

The Fock-space entropy operator under $\alpha$-partition is
\begin{equation}
    \alpha\mbox{-}\hat{S}_S := \hat{S}_{SS} + \alpha \hat{S}_{SR},
\end{equation}
and taking its quantum statistical average 
\begin{equation}
    \alpha\mbox{-}S^{(0)}_S = \mathrm{Tr}\{\hat{\rho}^{(0)} (\alpha\mbox{-}\hat{S}^{(0)}_S)\}
\end{equation}
yields  

\begin{eqnarray}\label{eqn_entropy_part_gce}
    \alpha\mbox{-}S^{(0)}_S(t) & = & 2\alpha \sum_\nu \langle \nu | \Pi_S |\nu\rangle s(\varepsilon_\nu) \nonumber \\
    &+&(1-2\alpha) \sum_{\mu,\nu} |\langle \mu |\Pi_S|\nu\rangle |^2 \nonumber \\
    & \times & [-f_\mu \ln f_\nu - (1-f_\mu)\ln(1-f_\nu)]\,.
\end{eqnarray}
Taking the limit of a continuous spectrum gives Eq.\ \eqref{eqn_entropy_part_open}.
For additional details, see Ref.\ \cite{webbHowPartitionQuantum2024}.

\section{Extension to multiple reservoirs}
\label{app_mult_resv}

In this section, we extend the analysis presented in the main text to the case of a multi-level system coupled to multiple reservoirs, each with the same
temperature and chemical potential (for the case of reservoirs with different temperatures and chemical potentials, see Refs.\ \cite{Shastry2020_STTh,staffordLocalTemperatureInteracting2016,shastryTemperatureVoltageMeasurement2016,shastryThirdLawThermodynamics2019,staffordLocalEntropyNonequilibrium2017,shastryTheoryThermodynamicMeasurements2019}).
The change in the grand potential of the $\alpha^{th}$ reservoir due to its coupling to the system can be written as
\begin{equation}
\delta\Omega^{(0)}_{R,{\alpha}}(t)=-\int_{-\infty}^{\infty} \,\frac{d\epsilon}{\pi} \mathbb{ImTr}\Bigg\{\Bigg(\frac{\partial \Sigma^{A}_\alpha(t,\epsilon)}{\partial \epsilon}\Bigg)\mathcal{G}^{A(0)}(t,\epsilon)\Bigg\}\omega(\epsilon)  \,,
\end{equation}
where $\Sigma^{A}_{nm,\alpha}(t,\epsilon)=\sum_kV_{kn,\alpha}(t)\mathcal{G}^{A}_{kk,\alpha}(\epsilon)V^{*}_{km,\alpha}(t)$ is the contribution to the advanced self-energy from the  $\alpha^{th}$ reservoir (with $\mathcal{G}^{A}_{kk,\alpha}(\epsilon)$ representing the advanced Green's function and $V_{kn,\alpha}$ representing the coupling element of the $n^{th}$ system orbital to the $k^{th}$ mode of the $\alpha^{th}$ reservoir, respectively)
so that its first variation is 
\begin{eqnarray}
   \dot{\Omega}_{R,\alpha}^{(1)}=&-&\int_{-\infty}^{\infty} \,\frac{d\epsilon}{\pi} \mathbb{ImTr}\Bigg\{\frac{\partial\Sigma^{A}_\alpha(t,\epsilon)}{\partial \epsilon}\frac{\partial \mathcal{G}^{A(0)}(t,\epsilon)}{\partial t}\Bigg\}\omega(\epsilon)\nonumber \\ 
  &-&\int_{-\infty}^{\infty} \,\frac{d\epsilon}{\pi} \mathbb{ImTr}\Bigg\{\frac{\partial^2\Sigma^{A}_\alpha(t,\epsilon)}{\partial t\partial \epsilon} \mathcal{G}^{A(0)}(t,\epsilon)\Bigg\}\omega(\epsilon)\,. \nonumber \\
 \end{eqnarray}
 Analogous expressions can be written for all other reservoir-partitioned First Law quantities. 

We illustrate this generalization to multiple reservoirs by analyzing the partitioned thermodynamics of a driven two-level system coupled to two reservoirs. The Hamiltonian for this model is given by $H(t)=H_{S}(t)+H_{R}+H_{SR}(t)$, where the system Hamiltonian is 
\begin{equation}
H_{S}(t)=\varepsilon_{1}(t)d_1^{\dagger}d_1 + \varepsilon_{2}d_2^{\dagger}d_2 + w (d_1^{\dagger}d_2 + \mbox{h.c.}) \,, 
\end{equation}
the two reservoirs are modeled as semi-infinite tight-binding chains with hopping integral $t_0$,
\begin{equation}
H_{R}= %
   \sum_{\alpha=1,2} \Bigg[t_0\sum_{j=1}^\infty (c_{j\alpha}^{\dagger}c_{j+1\alpha}+\mbox{h.c.})\Bigg] \,, 
\end{equation}
and the interface between system and reservoirs is modelled by 
\begin{equation}
H_{SR}(t)%
=\sum_{\alpha=1,2}V_{\alpha}(t)d_\alpha^{\dagger}c_{1\alpha}+\mbox{h.c.},
\end{equation}
wherein reservoir 1 is coupled to site 1 of the system and reservoir 2 is coupled to site 2 of the system.

\begin{figure}
\centering
\includegraphics{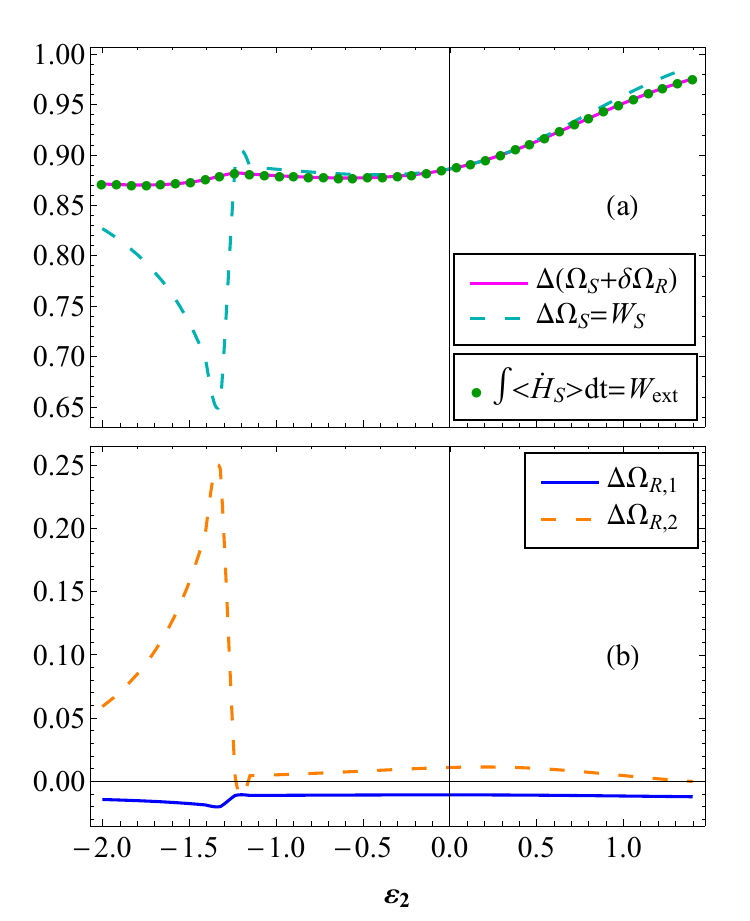} 
\captionsetup{justification=raggedright, singlelinecheck=false}	
\caption{%
Verification of the Work Sum rule [Eqs.\ \eqref{eqn_worksumrule_opensys}, \eqref{eqn_open_sum_rule_local}] for
an open quantum system: the two-level model coupled to two reservoirs. %
Only the level $\varepsilon_1(t)$ is driven from 0 to 1.5, and the coupling elements to the first and second reservoirs are fixed at $V_1=0.4$, $V_2=1.2$. Both reservoirs are maintained at temperature $T=0.02$ and chemical potential $\mu=1$, with  %
$t_0=1.25$, and the inter-level coupling is $w=0.5$. (a) Verification of the sum rule, (b) (Negative of) Nonlocal work resolved by reservoirs.
}
\label{fig_tls_sumrule_2res}
\end{figure}

Figure \ref{fig_tls_sumrule_2res} illustrates the work sum rule for the driven two-level system coupled to two reservoirs.  Only the energy level $\varepsilon_1(t)$ of site 1 is varied in the driving protocol, and the results are plotted versus the fixed energy level $\varepsilon_2$ of site 2. Fig.\ \ref{fig_tls_sumrule_2res} (a) illustrates the verification of the work sum rule, where the difference between  $W_S$ (dashed Cyan curve) and
$\Delta\Omega$ (solid Magenta curve) 
represents the total nonlocal quantum work. The nonlocal quantum work associated with the interface to reservoir 2, $-\Delta\Omega_{R,2}$ [Fig.\ \ref{fig_tls_sumrule_2res} (b)] is larger than that associated with the interface to reservoir 1, $-\Delta\Omega_{R,1}$ due to its stronger coupling to the system. The prominent feature as the fixed value of $\varepsilon_2$ approaches $-1.3$ from below in both Fig.\ \ref{fig_tls_sumrule_2res} (a) and (b) arises as the anti-bonding resonance of the two-level system is pushed up to and then above the chemical potential $\mu$ of the reservoirs.

\begin{figure}
    \centering
    \includegraphics[width=0.45\textwidth,height=15cm]{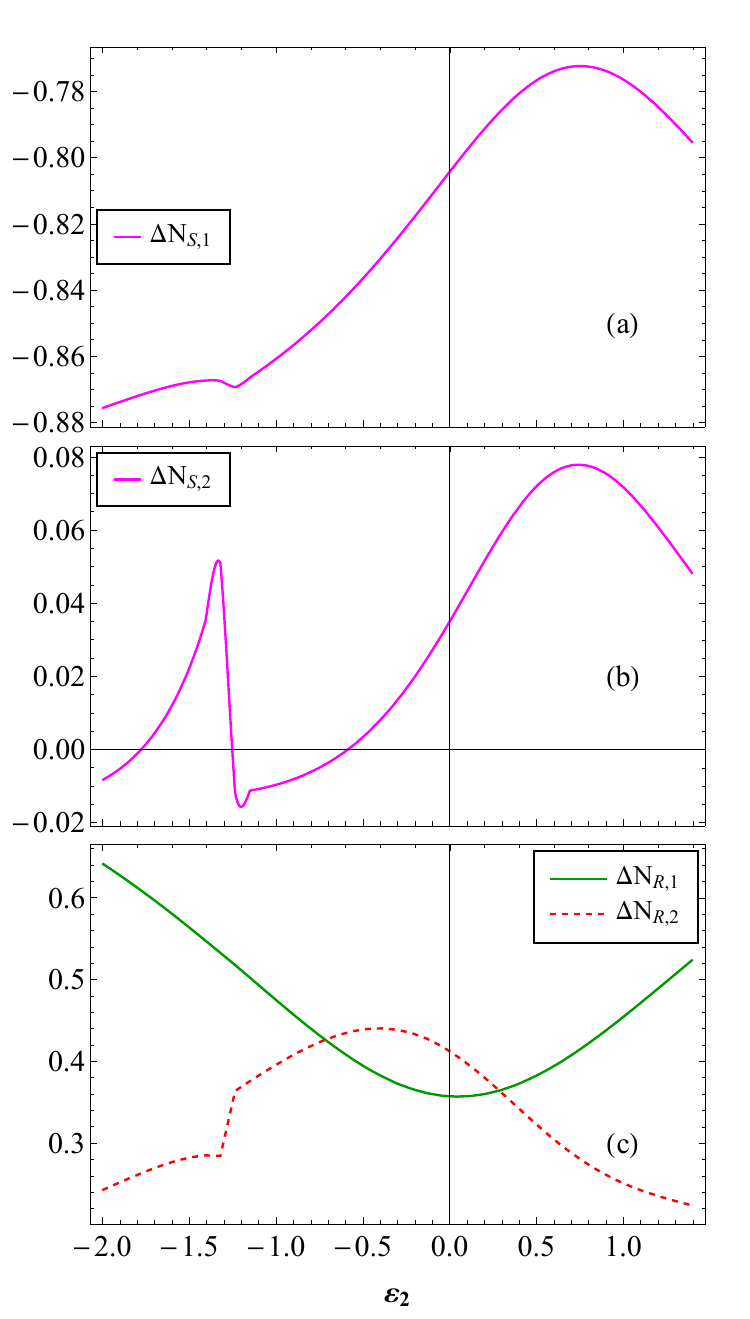}
    \captionsetup{justification=raggedright, singlelinecheck=false}
\caption{Spatially resolved change in particle number for the two-level model coupled to two reservoirs, displaying population-coherence coupled dynamics, where only the 
level $\varepsilon_1(t)$ is driven: (a) Change in particle number on the left site, (b) Change in particle number on the right site, and (c) Change in particle number in the left and right reservoirs. Here both the reservoirs have $T=0.02$ and $\mu=1$, while the level is varied from $\varepsilon_1$: $0\rightarrow 1.5$, and the coupling elements to the first and second reservoirs are fixed at $V_1=0.4$, $V_2=1.2$, and the inter-level coupling is $w=0.5$.}
\label{fig_tls_2res_partnum_grid}
\end{figure}

Figure \ref{fig_tls_2res_partnum_grid} shows the changes in the populations of the two sites of the system $\Delta N_{S,1}$ [Fig.\ \ref{fig_tls_2res_partnum_grid} (a)] and $\Delta N_{S,2}$ [Fig.\ \ref{fig_tls_2res_partnum_grid} (b)] and of the two reservoirs $\Delta N_{R,1}$ and $\Delta N_{R,2}$ [Fig.\ \ref{fig_tls_2res_partnum_grid} (c)] for the same driving protocol. Again, a sharp feature as the fixed value of $\varepsilon_2$ approaches $-1.3$ from below arises as particles are pushed out of the system when the anti-bonding resonance is pushed above $\mu$. This quantum machine functions as a coherent fermionic turnstile for $\varepsilon_2\in [-0.72, +0.28]$, as more particles are transferred to the second reservoir than the first during the protocol, although it is not a cyclic process, which could, however, be achieved trivially with an additional step in the protocol.

\begin{figure}
    \centering
    \includegraphics[width=0.45\textwidth,height=19cm]{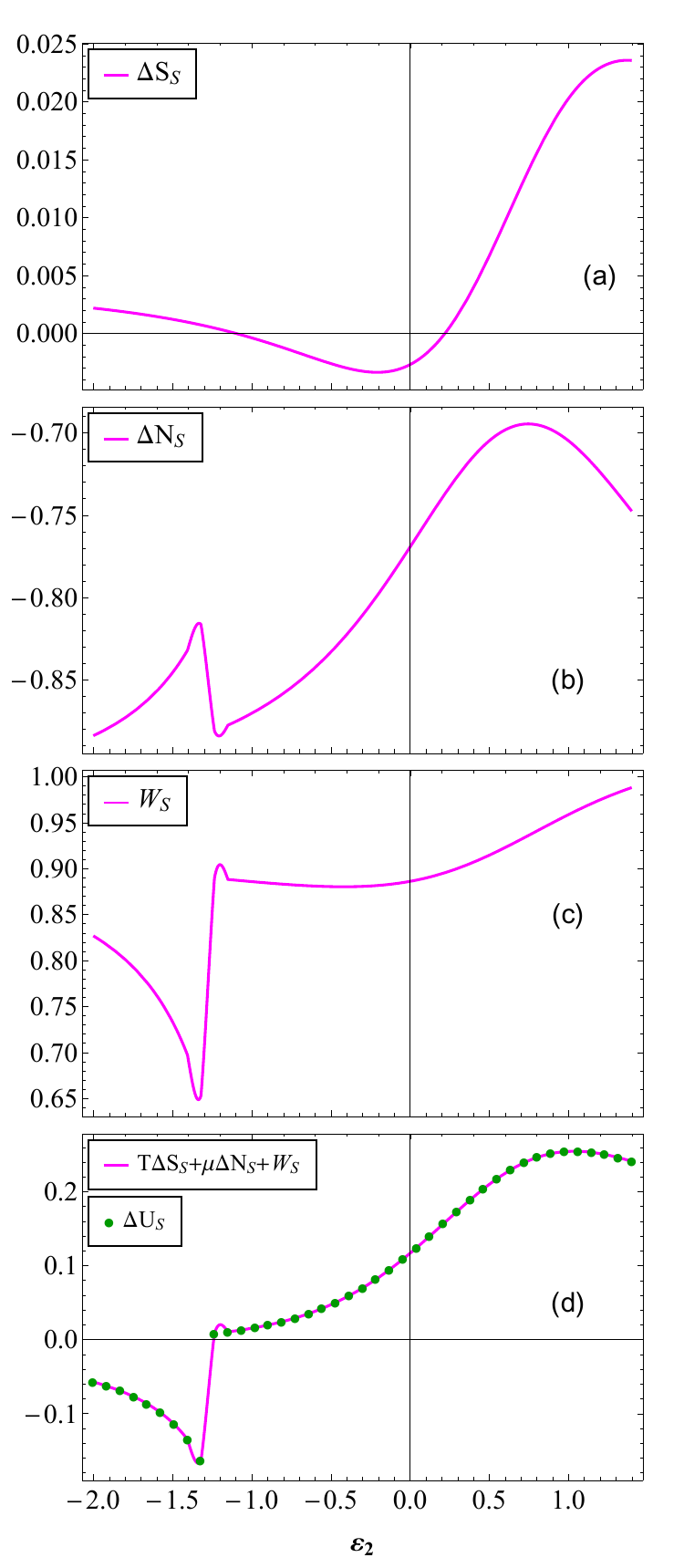}
    \captionsetup{justification=raggedright, singlelinecheck=false}
    \caption{The integrated terms (a) entropy $\Delta S_S$, (b) occupancy $\Delta N_S$, and (c) work $W_S$
    appearing in the 
First Law, Eq.\ \eqref{eqn_firstlaw_oqs}, %
 for the two-level model coupled to two reservoirs, where only the 
level $\varepsilon_1(t)$ is driven. (d) Verification of the equality of the LHS and RHS of Eq.\ \eqref{eqn_firstlaw_oqs}.  Here both the reservoirs have $T=0.02$ and $\mu=1$, while the level is varied from $\varepsilon_1$: $0\rightarrow 1.5$, and the coupling elements to the first and second reservoirs are fixed at $V_1=0.4$, $V_2=1.2$, and the inter-level coupling is $w=0.5$.}
\label{fig_tls_2res_firstlaw_grid}
\end{figure}

Finally, Fig.\ \ref{fig_tls_2res_firstlaw_grid} presents a verification of the First Law for the system under this driving protocol.

\end{document}